\documentclass[a4paper,12pt]{article}

\usepackage{graphics,graphicx,array}

\def \MET{\rm E{\!\!\!/}_T}
\usepackage{color}

\begin{document}

\begin{flushright} 
ADP-16-16/T971\\
KIAS-P16028
\end{flushright} 
\vspace{2.0cm}
\begin{center}
{\Large  \boldmath \bf
Di-Higgs signatures from R-parity violating supersymmetry as the origin of neutrino mass}
\\ \vspace*{0.2in} 
{\large Sanjoy Biswas$^1$, Eung Jin Chun$^1$ and Pankaj Sharma$^2$} \\
\vspace*{0.2in}
{\small \it
\small \it $^1$ Korea Institute for Advanced Study, Seoul 130-722, Korea\\
\small \it $^2$ Center of Excellence in Particle Physics (CoEPP),\\
\small \it The University of Adelaide, Adelaide, Australia}
\end{center}
\vspace*{0.6in}
\begin{center}
{\large\bf Abstract}
\end{center}

Motivated by the naturalness and neutrino mass generation, we study a bilinear R-parity violating supersymmetric scenario with a light Higgsino-like lightest supersymmetric particle (LSP). We observe that the LSP dominantly decays to $\nu h$ in a large part of the parameter space, and thus study the pair production of electroweakinos followed by the decays $\tilde\chi^\pm_1\to \tilde\chi^0_1 W^{\pm*}$ and $\tilde\chi^0_1\to \nu h$. This leads to an interesting signature of Higgs boson pair production associated with significantly large missing transverse energy which is grossly distinct
from the di-Higgs production in the Standard Model. We investigate the perspective of probing such signatures by performing a realistic detector level simulation of both the signal and corresponding backgrounds for the high-luminosity high energy phase of the Large Hadron Collider (LHC). We also advocate some observables based on kinematical features to provide an excellent handle to suppress the backgrounds.


\newpage
\section{Introduction}

Supersymmetry (SUSY) has been considered as the best framework to protect the electroweak scale from the quadratic divergence caused by a certain ultra-violet (UV) physics.  As a bonus, supersymmetry provides a good candidate of dark matter of the Universe: the lightest supersymmetric particle (LSP) which is neutral and stablized by assuming R-parity conservation. On the other hand, R-parity violation (RpV) brings another interesting possibility of generating tiny neutrino masses \cite{hall,oldies} in the context of the Minimal Supersymmegtric Standard Model (MSSM). The observed neutrino masses and mixings determine the lepton flavor structure of R-parity violating couplings which typically leads to clean signature of same-sign dileptons and predicts specific leptonic branching ratios of the LSP decay $\tilde{\chi}^0_1 \to l^\pm W^\mp$ \cite{mrv,jaja,valle2,chun02}. 

As no hint for supersymmetry appeared yet at the Large Hadron Collider (LHC), the naturalness argument for the TeV scale SUSY is in question due to a severe fine-tuning which turns out to be much more than expected. The electroweak symmetry breaking in SUSY requires a potential minimization condition:
\begin{equation} \label{mZmin}
{m_Z^2 \over 2} = {m^2_{H_d} - m^2_{H_u} \tan^2\beta \over \tan^2\beta -1 } - \mu^2
\end{equation}
where $m_{H_{u,d}}$ are the soft masses of the two Higgs doublets, $\tan\beta \equiv v_u/v_d$ is the ratio of their vacuum expectation values, and $\mu$ is the Higgs bilinear parameter in the superpotential. As the LHC pushes up the soft mass scale above TeV range, the condition (\ref{mZmin}) requires a fine cancellation among different terms. Barring too huge cancellation, one may arrange $m_{H_{u,d}}$ and  $\mu$ not too larger than $m_Z$ \cite{NS}, which still remains a viable option for SUSY.  It is a challenge for the LHC and future colliders to probe such a degenerate electroweakino \cite{ewkinos}.

In this paper, we investigate the LHC signatures of the light Higgsino in association with bilinear R-parity violation (BRpV) as the origin of the observed neutrino masses and mixings. Contrary to the conventional studies on BRpV predicting a peculiar signature of same-sign dileptons from $pp \to \tilde \chi_1^0 \tilde \chi_1^0  \to l^\pm l^\pm W^\mp W^\mp$, we will focus on the unusual case of the Higgsino LSP decay dominated by the Higgs channel $\tilde{\chi}^0_1 \to \nu h $ which will be shown to occur in a large region parameter space of the scenario under consideration.  As a consequence, it leads to an interesting LHC signature of  di-Higgs bosons with missing transverse energy. Measurement of Higgs-pair production cross-section will be one of the main focuses of the high energy and high luminosity LHC run. It is also an important step towards our understanding of the electroweak symmetry breaking mechanism. At LHC energies, Higgs boson pair production occurs dominantly through the gluon fusion  in the SM \cite{ggHH}. Other processes, such as weak boson fusion $qq^{(\prime)}\to qq^{(\prime)}hh$, associated productions $q\bar{q}^{(\prime)}\to Whh,~Zhh$ or associated production with top quarks, $gg,q\bar{q}\to t\bar t hh$ also occur albeit with cross sections which are 10-30 times smaller than the gluon fusion  \cite{Djouadi:1999rca,Gianotti:2002xx}.
Di-Higgs production at the LHC has been studied in the context of triliniear Higgs self coupling measurements by various authors \cite{SMHH} and references there in. While the SM cross sections are too small to allow observation at the LHC, it also plays a crucial role in the context of new physics searches at the LHC since physics beyond the SM can lead to an enhancement of the observable cross sections and/or different event kinematics. We demonstrate that di-Higgs in association with a large $\MET$ can be very useful to probe the BRpV SUSY where conventional channels fail to be sensitive.  Assuming only the electroweak production of the electroweakino pairs, we analyze the di-Higgs signal in the channel of $\gamma\gamma b \bar b \MET$ at the LHC14 with the integrated luminosity ranging from 1-3 ab$^{-1}$. Search for Higgs-pair production as a window to probe new physics is one of the major activities in the context of LHC and future collider, \textit{e.g.}, resonant Higgs-pair production in the context of singlet and doublet extension of the SM \cite{Dawson:2015haa,Hespel:2014sla}, double Higgs production via gluon fusion in the effective field theory framework \cite{Azatov:2015oxa}, Higgs pair production in the context of SUSY extension of the SM \cite{Cao:2013si,vanBeekveld:2015tka,Wu:2015nba} and various other extensions of it \cite{BSMHH}.

This paper is organized as follows: In Section 2 summarizing the results of Refs.~\cite{jaja, chun02}, we provide a brief review on the bilinear RpV couplings constrained by  the resulting tree-level neutrino mass matrix. In Section 3, the Higgsino-like LSP decay modes are analyzed for the cases of $\mu < M_1 \leq M_2$ where $M_{1,2}$ are the masses of the bino and wino components, respectively. In Section 4, we compute the  Higgsino pair production cross-section for some benchmark points and perform a realistic detector-level simulation for signals and backgrounds in the di-Higgs decay channel of $hh \to \gamma\gamma b \bar b$ and obtain the LHC14 perspective to probe our scenario. Finally, we summarize our results and conclude in Section 5. In Appendix A, we collect the effective R-parity violating couplings relevant for the Higgsino decays and  Appendix B shows the decay widths of the neutralinos induced by the BRpV couplings.

\section{Bilinear RpV and neutrino mass matrix}

Allowing lepton number violation in the supersymmetric standard model, the superpotential is composed of the  R-parity conserving $W_0$ and violating $W_1$ part;
\begin{eqnarray} \label{supo}
 W_0 &= & \mu H_1 H_2 + h^e_{ij} L_i H_2 E^c_j + h^d_{ij} Q_i H_2 D^c_j 
         + h^u_{ij} Q_i H_1 U^c_j  \nonumber\\
 W_1 &=& \epsilon_i \mu L_i H_1 +  {1\over2}\lambda_{ijk} L_i L_j E^c_k
      +    \lambda^\prime_{ijk} L_i Q_j D^c_k \,.      
\end{eqnarray}
Among soft supersymmetry breaking terms, let us write R-parity 
violating bilinear terms;
\begin{equation}
 V_{soft} = B\mu H_1 H_2 + B_i \epsilon_i \mu  L_i H_1  
            +  m^2_{L_iH_2} L_i H_2^\dagger + h.c. \,.
\end{equation}
It is clear that the electroweak symmetry breaking gives rise 
to nonzero vacuum expectation values of sneutrino fields, $\tilde{\nu}_i$, 
as follows:
\begin{equation}
 a_i \equiv {\langle \tilde{\nu}_i \rangle \over 
   \langle H_2 \rangle} =  
     - { \bar{m}^2_{L_iH_2} + B_i \epsilon_i \mu t_\beta 
             \over m^2_{\tilde{\nu}_i} }
\end{equation}
where $\bar{m}^2_{L_iH_2}= m^2_{L_iH_2}+ \epsilon_i \mu^2$, 
$t_\beta=\tan\beta = \langle H_1 \rangle/ \langle H_2 \rangle$ and 
$m^2_{\tilde{\nu}_i}= m^2_{L_i}+ M_Z^2 c_{2\beta}/2$.

\medskip

Given the BRpV couplings $\epsilon_i$ and $a_i$, the neutrino-neutralino sector form a $7\times7$ mass matrix whose $3\times4$ (Dirac) neutrino-neutralino mass matrix takes the form of 
\begin{equation} \label{MD}
{\cal M}^D_{ij} = ( - a_i c_\beta M_Z s_W, a_i c_\beta M_Z c_W, 0, \epsilon_i \mu)
\end{equation}
where $s_W\equiv \sin\theta_W$ is the weak mixing angle, and the index $i$ runs for three neutrino flavors $(\nu_e, \nu_\mu, \nu_\tau)$, and $j$ runs for the neutralino states ($\tilde{B}$, $\tilde{W}_3$, $\tilde{H}^0_1$, $\tilde{H}^0_2$) which has the usual $4\times4$ mass matrix ${\cal M}^N$ containing the bino, wino and Higgsino masses denoted by $M_1$, $M_2$ and $\mu$, respectively.

As is well-known, a seesaw diagonalization rotating away ${\cal M}^D$ [see Appendix A for details] generates the ``tree-level''  neutrino mass matrix ${\cal M}^\nu = -{\cal M}^D {{\cal M}^N}^{-1} 
{{\cal M}^{D}}^T$ whose components are given by
\begin{equation} \label{Mtree}
 {\cal M}^{\nu}_{ij}= - {M_Z^2 \over F_N} \xi_i \xi_j c_\beta^2 \,,
\end{equation}
where $\xi_i \equiv a_i -\epsilon_i$ and  $F_N=M_1 M_2 /( c_W^2 M_1 + s_W^2 M_2) + M_Z^2 s_{2\beta}/\mu$. This makes massive only one neutrino, $\nu_3$, in the direction of $\vec{\xi}$.  The other two get masses from finite one-loop corrections and thus $\nu_3$ is usually the heaviest component.  We fix the value of $m_{\nu_3}$  from the atmospheric neutrino data  and thus the overall size of $\xi\equiv |\vec{\xi}|$ is determined to be
\begin{equation} \label{xicb}
 \xi c_\beta = 0.74\times10^{-6} \left(F_N \over M_Z\right)^{1/2}
               \left(m_{\nu_3} \over 0.05 \mbox{ eV} \right)^{1/2}\,.
\end{equation}
Furthermore, among three neutrino mixing angles defined by the mixing matrix 
\begin{equation}
 U=\pmatrix{ 1 & 0 & 0 \cr 0 & c_{23} & s_{23} \cr 0 & -s_{23} & c_{23} \cr}
   \pmatrix{ c_{13} & 0 & s_{13} \cr 0 & 1 & 0 \cr -s_{13} & 0 & c_{13} \cr}
   \pmatrix{ c_{12} & s_{12} & 0 \cr -s_{12} & c_{12} & 0 \cr 0 & 0 & 1 \cr}
\end{equation}
with $c_{ij}=\cos\theta_{ij}$ and $s_{ij}=\sin\theta_{ij}$, etc., two angles are  determined by the tree-level mass matrix (\ref{Mtree}) as follows:
\begin{eqnarray} \label{twoangles}
  \sin^22\theta_{23} &\approx& 4 {\xi_2^2 \over \xi^2}
                                 {\xi_3^2\over \xi^2} \nonumber\\
  \sin^22\theta_{13} &\approx& 4 {\xi_1^2 \over \xi^2}
                         \left(1-{\xi_1^2\over \xi^2}\right)  \,.
\end{eqnarray}
These two angles define the atmospheric and reactor neutrino oscillation angles, respectively, and thus one has $\sin^22\theta_{23} \approx 1$ and $\sin^22\theta_{13} \approx 0.09$ \cite{pdg}. 
This implies that the sizes of $\xi_i$ should follow the relation:
\begin{equation} \label{xi_values}
|\xi_1| : |\xi_2| : |\xi_3| \approx 0.1 : 1 : 1 \,.
\end{equation} 
The other angle $\theta_{12}$ can be determined only after including one-loop corrections which are assumed to be smaller than the tree-level contribution (\ref{Mtree}) and thus irrelevant for our discussion. 

The BRpV terms induce mixing between  neutrinos (charged leptons) and neutralino (charginos) as well as their scalar partners. Rotating them away, one gets the effective RpV vertices of neutralinos and charginos which are summarized in Appendix A.  

\section{Light Higgsino Decays}

A distinct feature of the RpV SUSY models is that the LSP, $\tilde\chi^0_1$,  is not  stable. In our analysis,  neutralino decays via sfermions are highly suppressed  in the limit of heavy sfermion masses and small trilinear RpV couplings responsible for one-loop neutrino mass generation. \begin{figure}[h!]
\centering
\includegraphics[width=\textwidth]{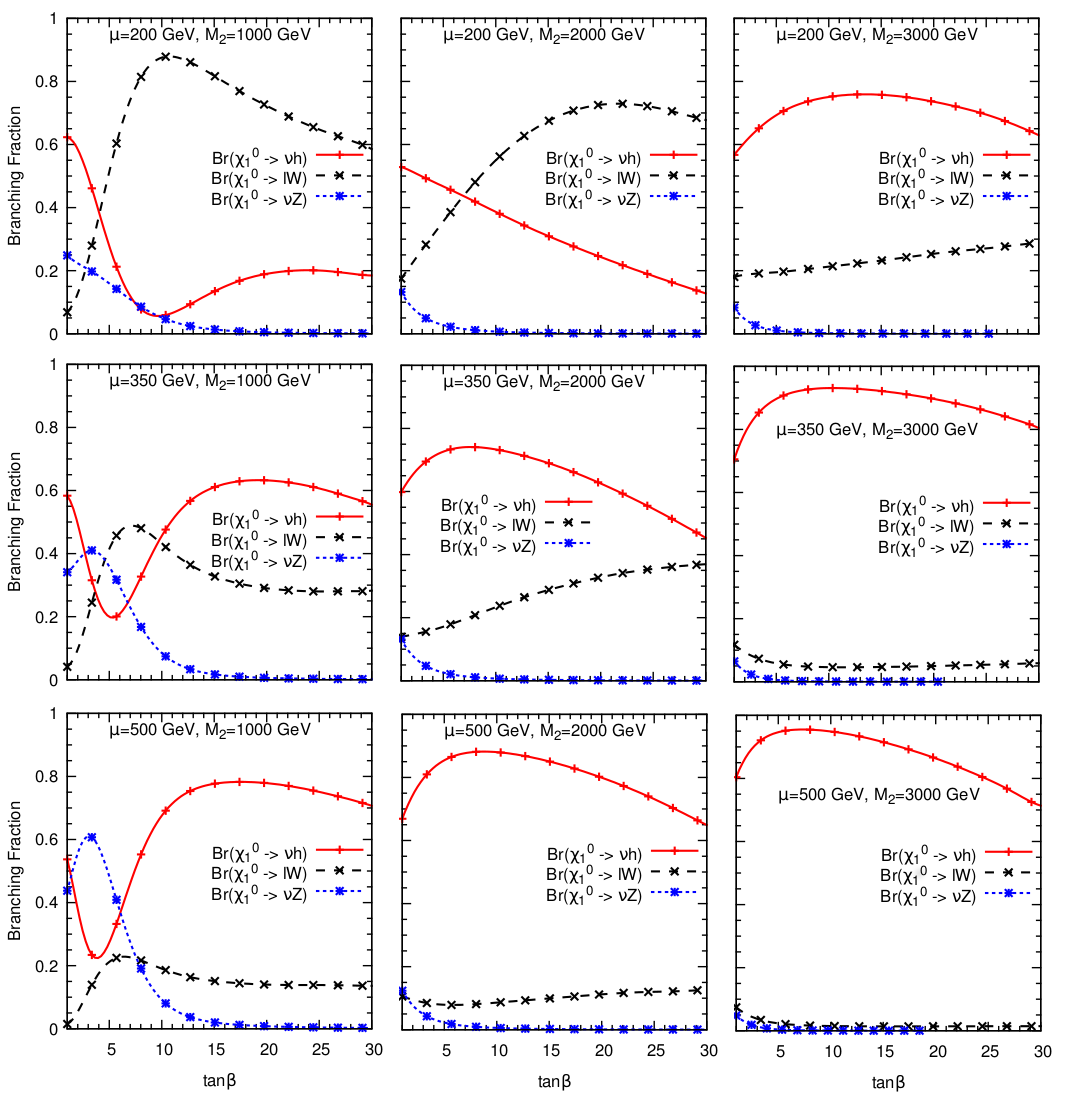}
\caption{Branching fractions for the decays $\tilde\chi_1^0\to \nu h$, $\tilde\chi_1^0\to \ell W$ and $\tilde\chi_1^0 \to \nu Z$ in terms of $\tan\beta$ for different values of $\mu$ and $M_2$. \label{fig:br_chi0}}
\end{figure}
Therefore, we discuss the branching ratios (BR) of the LSP for the following decay processes:
\begin{equation}
 \tilde\chi_1^0\to Z\nu_\ell,~~~~~~\tilde\chi_1^0\to l W,~~~~~~\tilde\chi_1^0\to h\nu_\ell
\end{equation}
where $\ell$ is any of the three charged-leptons.

To explore the phenomenological features, we consider a specific scenario in this work where $\mu<M_1, M_2$ leading to a situation of the Higgsino LSP, with the lightest states $\tilde\chi_{1,2}^0$ and $\tilde\chi_1^\pm$ being Higgsino-like. If $M_1<M_2$, $\tilde\chi^0_3$ is bino-like while $\tilde\chi^\pm_2$ and $\tilde\chi^0_4$ are wino-like; and for $M_2<M_1$, $\tilde\chi^0_4$ is bino-like while $\tilde\chi^\pm_2$ and $\tilde\chi^0_3$ are wino-like.

In the following, we fix $M_1=1.0$ TeV and vary both $\mu$ and $M_2$. Given $M_1$ and $M_2$, the masses of electroweakinos are determined by the value of $\mu$. We also vary $\tan\beta$ to see the variation of branching ratios with respect to $\mu$, $M_2$ and $\tan\beta$ (Fig.~\ref{fig:br_chi0}). Fig. \ref{fig:br_chi0}  shows that the branching ratio for the $\ell W$ or $\nu h$ decay mode is quite sensitive to the values of $\mu$, $M_2$ and $\tan\beta$.

All of the RpV vertices which we have obtained in Eqs.~\ref{Vert:chilW}-\ref{Vert:chalh} of Appendix A depend only on one type of BRpV variables $\xi_i$ which have taken to satisfy $\xi_1:\xi_2:\xi_3 = 0.1:1:1$ (Eq.~\ref{xi_values}). Thus,  BR($\tilde\chi_1^0\to eW$) is quite suppressed and major contribution to $lW$ mode comes from $\mu W$ and $\tau W$ channels. An important point to note is that the decay rate ($\tilde\chi_1^0\to \ell W$) also depends on charged-lepton masses through $c_2^R$ (Eq.~\ref{width:lW}). Although this mass dependence is suppressed by $m^\ell/F_C$, its contribution can be substantial for $\tau$-lepton at large $\tan\beta$ as $c_2^R$ is proportional to $\tan\beta$ (Eq.~\ref{cR}). Thus, BR($\tilde\chi_1^0\to \ell W$)  grows at large $\tan\beta$ through the enhancement of $\tau W$ decay width. 

Another important aspect to note from Eq.~\ref{cR} is that both $c_i^L$ and $c_i^R$ are inversely proportional to $\mu$. Thus for small values of $\mu$, the decay rate $\Gamma(\tilde\chi_1^0\to \ell W)$ is much larger than $\nu Z$ and $\nu h$ decay modes. This feature can be seen from the top row of Fig.~\ref{fig:br_chi0} where we have displayed the BRs of $\ell W$, $\nu h$ and $\nu Z$ decay modes for $\mu=200$ GeV. In these figures, it can be seen that BR($\ell W$) dominates at large $\tan\beta$. Decay mode $\nu h$ is only significant at low $\tan\beta$. For larger value of $\mu=500$ GeV (see bottom panel of Fig.~\ref{fig:br_chi0}), BR($\ell W$) becomes subdominant as expected. Thus, the dominant decay channel with the branching ratio varying between (0.6-0.9) is the $\nu h$ mode for all values of $\tan\beta$. BR($\nu h$) decreases by about 20\% because of the increase in the decay rate of $\chi_1^0\to \ell W$ at large $\tan\beta$. 

\begin{figure}[h!]
\includegraphics[width=\textwidth]{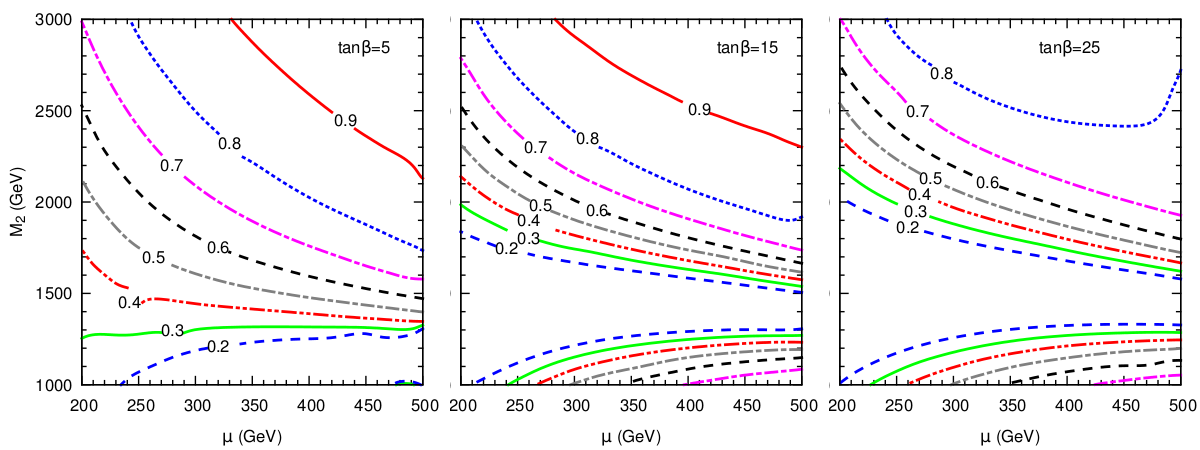}
\caption{\label{fig:mu-m2} Branching ratio of the decay $\tilde\chi_1^0 \to \nu h$ in the $M_2-\mu$ plane for three different values of $\tan\beta= $ 5 (left), 15 (center) and 25 (right).}
\end{figure}

From the Fig. \ref{fig:br_chi0}, we conclude that for large $M_2=3$ TeV, the BR($\tilde\chi_1^0\to \nu h$) always dominates over all other $\tilde\chi^0_1$ decays. For moderate $M_2=2$ TeV, the BR($\tilde\chi_1^0\to \nu h$) still dominates over $\ell W$ and $\nu Z$ modes but for low values of $\tan\beta$ while for small $M_2=1$ TeV, we find that BR($\tilde\chi_1^0\to \nu h$) dominates but only for large value of $\tan\beta$ and for large values of $\mu$. All these facts can easily be deduced from the effective RpV couplings which are derived in Appendix A. To have better understanding of the behavior of BR($\tilde\chi_1^0\to \nu h$), we plot the BR in the plane of $(M_2,\mu)$ for three different values of $\tan\beta=$ 5 (left), 15 (center) and 25 (right) in Fig. \ref{fig:mu-m2}. One can see from the Fig. that the largest possible values of the BR($\tilde\chi_1^0\to \nu h$) can be achieved when $\tan\beta$ is small and both $M_2$ and $\mu$ is large.

\section{Light Higgsino production at the LHC}

\begin{figure}[h!]\centering
\includegraphics[scale=0.95]{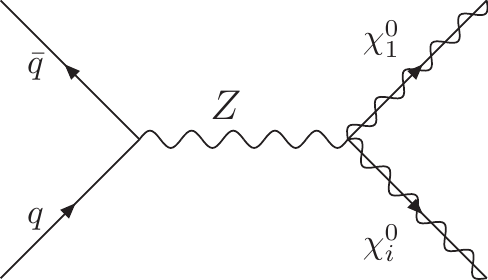}\hskip 25pt
\includegraphics[scale=0.95]{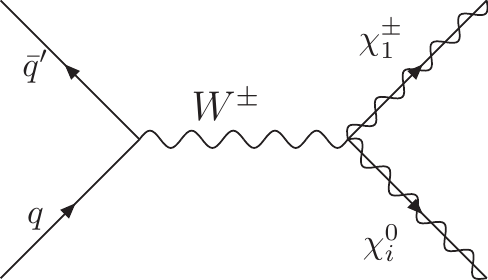}
\caption{\label{feynm-diag}Feynman diagrams representing pair production of electroweakinos. Similar diagram for chargino pair production mediated via $\gamma^*$ and $Z$ boson also exist.}
\end{figure}

\begin{figure}[h!]\centering
\includegraphics[scale=0.75]{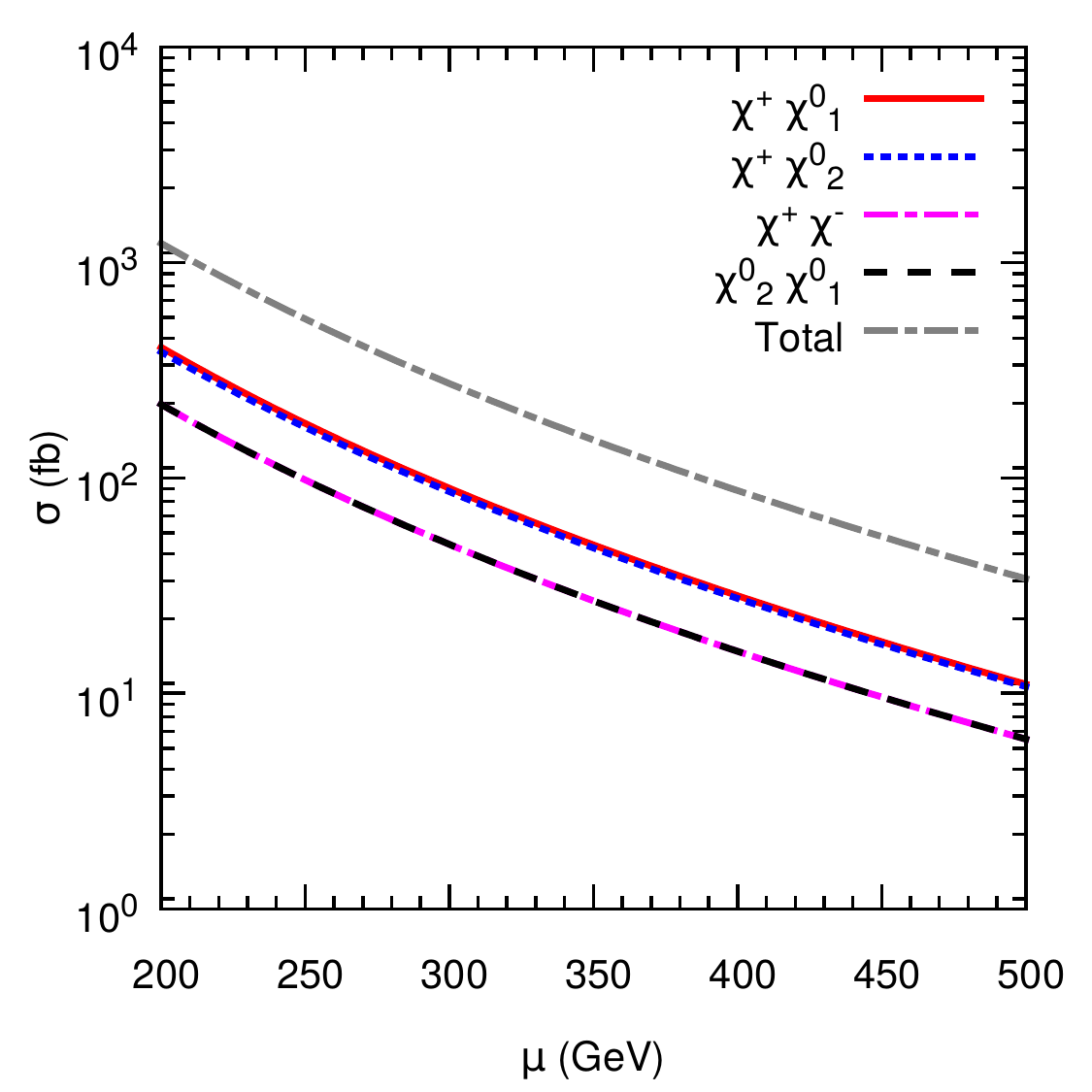}
\caption{\label{xsec-mu}Production cross sections of the charged and neutral Higgsinos in terms of $\mu$ at the LHC14.}
\end{figure}

From the analysis of previous section, we conclude that the decay channel $\tilde\chi_1^0 \to \nu h$ is significant in large part of the parameter space and, in fact, close to $\sim90-95$\% in certain regions. Motivated by this observation, we consider a very interesting signature of Higgs boson pair-production at the LHC in RpV SUSY. To probe this yet unexplored parameter space through di-Higgs, we consider pair production of Higgsinos at the 14 TeV LHC run:

\begin{enumerate}
 \item Neutralino pair production: $pp \to \tilde\chi_1^0 \tilde\chi_i^{0}$,
 \item Chargino pair production : $pp \to \tilde\chi^+_1 \tilde\chi_1^-$, 
 \item Associated neutralino and chargino production : $pp \to \tilde\chi_1^\pm \tilde\chi_i^0$,
\end{enumerate}
where $i=1,2$. The main contributions to these processes come from $s$-channel mediation of $\gamma$, Z and $W^\pm$ bosons (see Fig. \ref{feynm-diag}). The contributions from $t$-channel squark mediated processes are suppressed by heavy squark masses and can be ignored. Hence, the cross sections only depend on the masses of the electroweak gauginos and their couplings with $\gamma$, Z and $W^\pm$ bosons. We have worked with leading order (LO) cross-sections. The effects of next-to-leading order (NLO) QCD corrections can be achieved by including multiplicative K factors which can give a 10-20\% enhancement. However, here we present a conservative estimate without multiplying by any K-factors. In Fig.~\ref{xsec-mu}, we show the cross sections for these production channels as a function of $\mu$ at 14 TeV LHC. The cross sections for the associated production are the largest followed by the chargino and neutralino pair productions.

The chargino mainly decays via R-parity conserving coupling to $\tilde\chi_1^\pm\to \tilde\chi_1^0 W^{\pm *}\to ff^{\prime}\tilde\chi_1^0$ almost 90\% of time in all regions of the parameter space. The R-parity violating decays ($\ell h, \nu W, \ell Z$) are suppressed compared to the R-parity conserving one due to the tiny RPV couplings. Thus, in most of the events, one ends up with a pair of LSP's which subsequently decays to $\tilde{\chi}_1^0\to \nu h$ via R-parity violating interactions. This naturally leads to pair production of Higgs and two invisible neutrinos which contribute to missing transverse energy ($\MET$). The signal we investigate at the LHC, therefore, consists of $pp\to h h + \MET + X$, where $X$ stands for additional jets and/or unclustered particles but no isolated leptons.

 \begin{table}[h]
 \centering
 \newcolumntype{C}[1]{>{\centering\let\newline\\\arraybackslash\hspace{0pt}}m{#1}} 
\begin{tabular}{ ||C{4.5cm}|C{3.cm}||}  \hline
  Channel & BR(\%)\\
  \hline\hline
  bbbb		&	36\\
  bbWW 		& 	24.7\\
  $bb\tau\tau$	&	7.3\\
  $WWWW$	& 	4.3\\
  $bb\gamma\gamma$	& 0.27\\
  $bbZZ(\to e^+e^-\mu^+\mu^-)$ & 0.015 \\
  $\gamma\gamma\gamma\gamma$   & 0.00052\\
  \hline
  \end{tabular}
\caption{\label{BR:Higgs}Branching ratios for different di-Higgs channels.}
 \end{table}

In Table \ref{BR:Higgs}, we show the branching ratios of di-Higgs decays in various channels. The dominant decay of the SM-like Higss to a pair of $b$ quarks occurs with a branching ratio of 60\%. Thus, the dominant branching ratio for di-Higgs decay is to $hh\to bbbb$, which is $\sim 36\%$. But the backgrounds for 4$b$ and $2b2\tau$ final states are overwhelming. While the backgrounds for $4\tau$ decay mode are moderate, it suffers from small $\tau$ detection efficiencies. As far as decays to electroweak gauge bosons are concerned, $bbWW$ and $bbZZ$ decaying to $b\bar b\ell^+\ell^-\nu \bar\nu$ suffer from huge QCD $t\bar t$ pair backgrounds. The $b\bar b ZZ^*\to b \bar b+4$ leptons and $b\bar b Z\gamma$ channels suffer from too small rates. So, this leaves us a very few options to choose from. Among all the di-Higgs decay channels, we focus on the possibility where one Higgs decays to a $b\bar b$ pair and the other to a di-photon pair.  Thus the final signature for di-Higgs search of interest in this work is $\gamma\gamma b \bar b \MET +X$ at the LHC. 

We generate the spectrum for the BRpV SUSY model using {\tt SARAH} \cite{Staub:2008uz} and {\tt SPheno} \cite{spheno}, and then, use these spectrum files in the {\tt PYTHIA} \cite{pythia6} to generate the events. We also use {\tt PYTHIA} for the parton showering and hadronisation. We have used simple cone algorithm of {\tt PYCELL} inside {\tt PYTHIA} to form jets out of clusters of hadrons with a cone size of $R=0.4$. For a realistic detector simulation, we also include the  appropriate Gaussian smearing of the energies of each objects (i.e., jets and photon) in an event using the following resolution function, $$\frac{\Delta E}{E}=\frac{a}{E}\oplus\frac{b}{\sqrt{E}}\oplus c.$$ For jets, we take $a = 1$ GeV,~ $b = 0.8$ GeV$^{1/2}$,~ $c = 0.05$ while for photons the values are $a = 0.35$ GeV,~ $b = 0.07$ GeV$^{1/2}$,~ $c = 0.007$ \cite{jet-resl,ele-pho-resl}. Here, E is in the units of GeV. The identification efficiencies for a true $b$-jet, photon and their respective mistagging  probabilities have been given in Table \ref{eff}. The tagging and mistagging efficiencies are more or less consistent with the ATLAS experiment and have been taken from \cite{ATL-eff}.

In the following, we will perform a detailed collider simulation for the di-Higgs production process and the corresponding backgrounds at the 14 TeV high luminosity LHC with the luminosity of 3 ab$^{-1}$ in the BRpV model. We will also discuss the strategy and cuts required to suppress the backgrounds and enhance the signal-to-background ratio. In our simulation, we consider different values of Higgsino masses, ranging from 200 GeV to 600 GeV, for the signal reconstruction and analysis. However for the purpose of illustration, we consider the following benchmark point for the model where BR($\tilde\chi^0_1\to \nu h$) is maximized: $\mu=300$ GeV, $\tan\beta=15$, $M_2=3$ TeV. For this benchmark point, we have $M_{\tilde\chi^0_1}=300$ GeV, $M_{\tilde\chi^\pm_1}=310$ GeV, BR$(\tilde\chi^0_1\to \nu h)=0.9$, BR$(\tilde\chi_1^+\to \tilde\chi^0_1 W^+)=0.95$ and $\sigma(pp\to\chi\chi)=232$ fb. We have shown all the figures for the various distribution for this benchmark point. Nevertheless we also present our analysis for various Higgsino masses in Table \ref{cut-flow:sig}.

\begin{table}
\centering
 \newcolumntype{C}[1]{>{\centering\let\newline\\\arraybackslash\hspace{0pt}}m{#1}} 
\begin{tabular}{ C{1.5cm}C{1.5cm}C{1.5cm} C{1.5cm}  C{1.5cm} C{1.5cm} }

 \hline
 \hline
 $\epsilon_\gamma$	& $\epsilon_b$	& $P_{c\to b}$	& $P_{\tau\to b}$ & $P_{j\to b}$	& $P_{j \to \gamma}$ \\
 \hline
 90\%	& 70\%	& $1/8$	& 1/26 & $1/440$ & $1/1000$ \\
 \hline
 \end{tabular}
\caption{\label{eff} Photon and $b$ jet identification efficiencies and misidentification probabilities for charm quarks, $\tau$, light jets to $b$ jets and photons at the LHC. \cite{ATL-eff}}
\end{table}

\subsection{$\gamma\gamma b \bar b$ Channel}
\label{cuts}

We perform the signal calculation $pp\to hh \ \MET + X\to b\bar b\gamma\gamma \ \MET+X$.
Though the branching ratio of the $\gamma\gamma b \bar b$ channel is very small, it can reach the similar sensitivity of the $b\bar b WW$ channel in probing the di-Higgs signals because of the precise resolution of two photons which leads to a very prominent peak around Higgs mass in $M_{\gamma\gamma}$ distribution. The background processes for the $\gamma\gamma b\bar b$ channel are QCD $b\bar b\gamma\gamma$, $b\bar b h(\to \gamma\gamma)$, $\gamma\gamma h(\to b\bar b)$ and multijet QCD backgrounds resulting from jets faking either as $b$-jets or photons like $jj\gamma\gamma$ with two fake $b$ jets; $b\bar b j\gamma$ with $j$ faking photon; $b\bar b jj$ with two fake photons; $jjj\gamma$ with two fake $b$ jets and one fake photon; $jjjj$ with two fake $b$-jets and two photons; $hjj$ with either two fake $b$ jets or two fake photons; and $hj\gamma$ with one fake photon. While estimating and generating multijet backgrounds, we consider the misidentified charm quarks separately from the light flavour jets because of the very different mistagging factors as given in Table \ref{eff}. 

In the following we present our strategy to suppress the background and enhance the signal-to-background ratio. The acceptance and selection cuts applied in our analysis are as follows:

\begin{itemize}
 \item {\bf Identification cuts (Cut 1):}
 \begin{enumerate}
  \item Accept events with two photons, 2 $b$-jets and missing energy,
  \item Photons must have transverse momentum $p_T^\gamma>10$ GeV and rapidity $|\eta^\ell|<2.5$,
  \item All $b$-jets must have following $p_T$ and $\eta$ requirements:
  $$p_{T}^b>20~ \mbox{GeV}, |\eta^b|<2.5$$
  \item All pairs of jets, photons and photon plus jets should be well separated with each other by:
  $$\Delta R_{jj,jb,bb,\gamma j,\gamma b,\gamma\gamma}\geq 0.4~~ \mbox{where}~~\Delta R=\sqrt{(\Delta \phi)^2+(\Delta \eta)^2}$$
 \end{enumerate}
\item {\bf Selection requirements:} When an event satisfies above requirements, it is further processed for the 
signal reconstruction and background reduction as follows:

\begin{figure}[h!]
\includegraphics[scale=0.65]{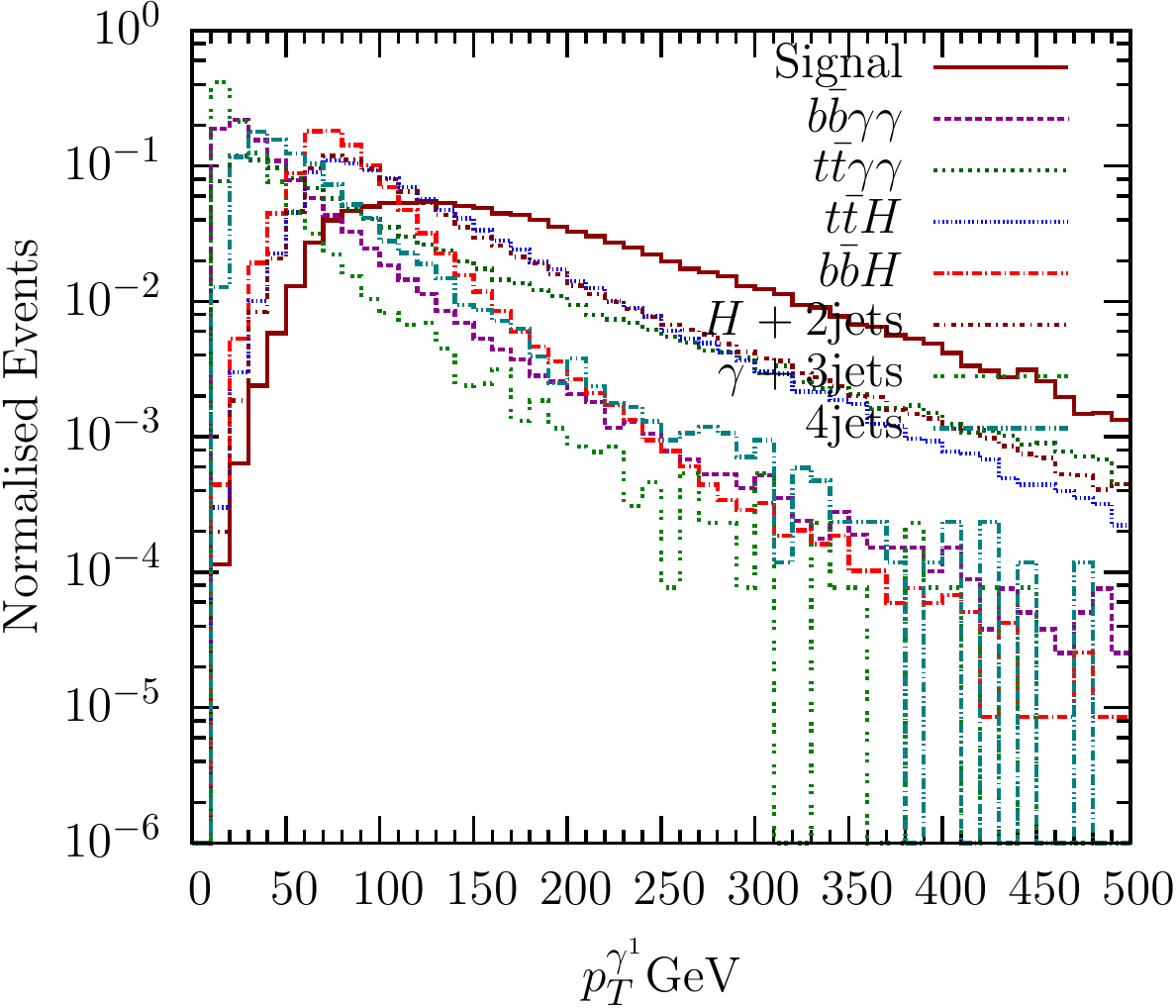}\hspace{5mm}
\includegraphics[scale=0.65]{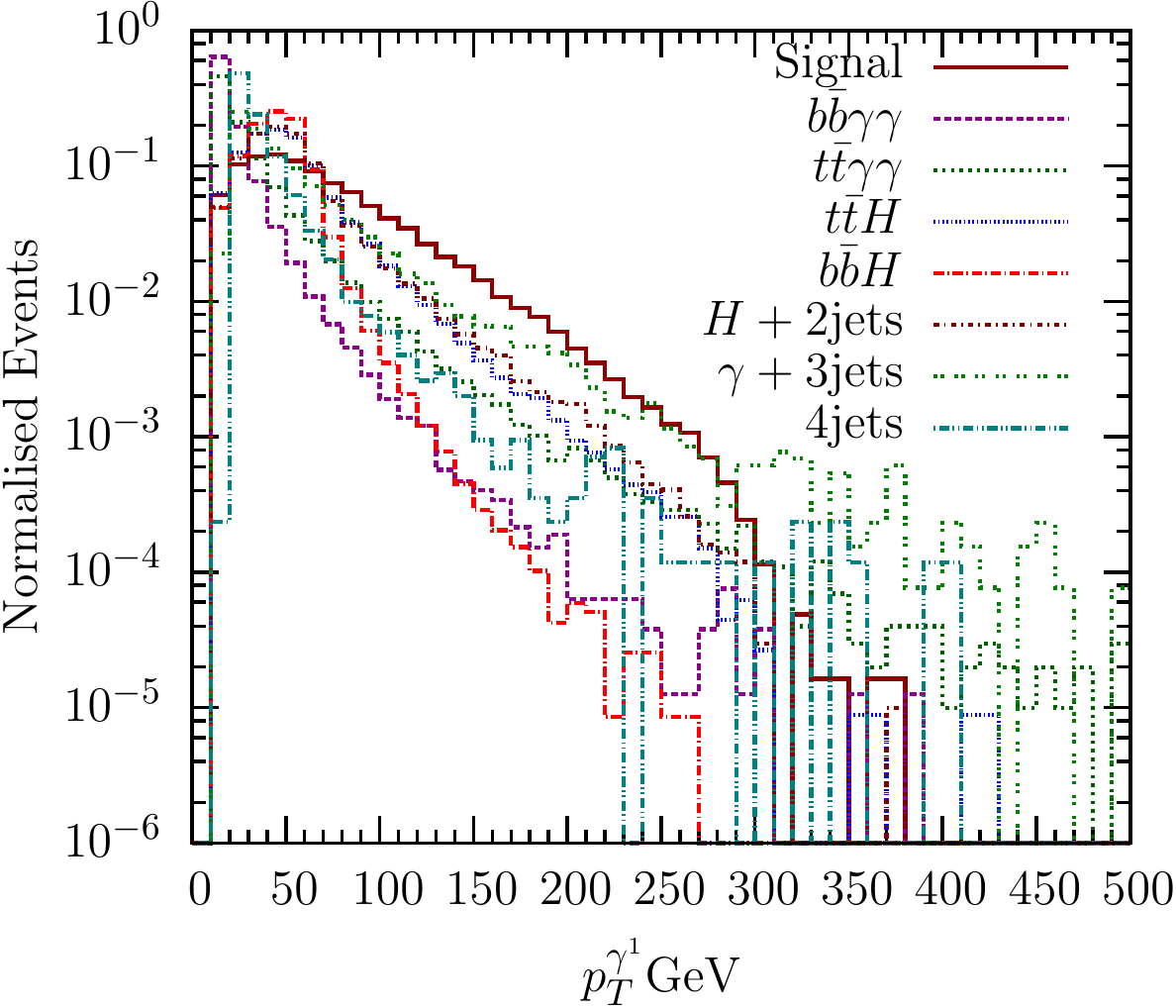}
\caption{\label{bbgg:pT_photon}
Transverse momentum distribution of the two photons for the signal and backgrounds.}
\end{figure}

\begin{itemize}
 \item {\bf Cut on $p_T$ of photons (Cut 2)}: To get rid of soft photons coming from the decay of mesons or radiations, we further put the following $p_T$ cuts on the two photons: 
 $$p_{T}^{\gamma_1}>30 \mbox{GeV} \quad\mbox{and}\quad p_{T}^{\gamma_2}>20 \mbox{GeV}. $$ In Fig.~\ref{bbgg:pT_photon}, we show the transverse momentum distribution of the two photons for the signal as well as for the backgrounds before the cuts. For the hardest photon, the distribution peaks at around 100 GeV while most backgrounds peak at low $p_T$. 

\item {\bf Invariant mass cuts (Cut 3)}: In Fig.~\ref{bbgg:InvM}, we show the invariant mass distribution of two photons, $M_{\gamma\gamma}$ and two $b$ jets, $M_{b\bar b}$ for both the signal and backgrounds. In the case of the signal and some of the backgrounds, the two photons come from a resonant Higgs. Thus, the diphoton invariant mass distributions for the resonant Higgs have a very narrow peak around $M_h=125$ GeV. On the other hand,  for other backgrounds, the distribution is continuum with no localised events around $M_h$. Also, the fact that the photons are measured at a very high degree of precision at the LHC, allows a very sharp distribution even at the detector level. On the other hand, due to the large jet energy uncertainties and energy resolution of the jets, the invariant mass distributions of the resonant Higgs decaying to two $b$-jets show a relatively-wider peak. The peak is also shifted towards lower value (around 112.5 GeV) than the actual Higgs mass due to invisible neutrinos coming from $b$-decays and missing particles outside the jet cone. Both the diphoton and di-$b$-jet invariant mass distributions are quite distinct from the background and have  prominent peaks. We utilize this to suppress the backgrounds and further cuts on invariant masses are imposed:
$$|M_{\gamma\gamma}-M_h|<2.5~\mbox{GeV},~~~~~~~~~~~~~|M_{b\bar b}-112.5|<15~\mbox{GeV},$$ 

\begin{figure}[h!]
\includegraphics[scale=0.65]{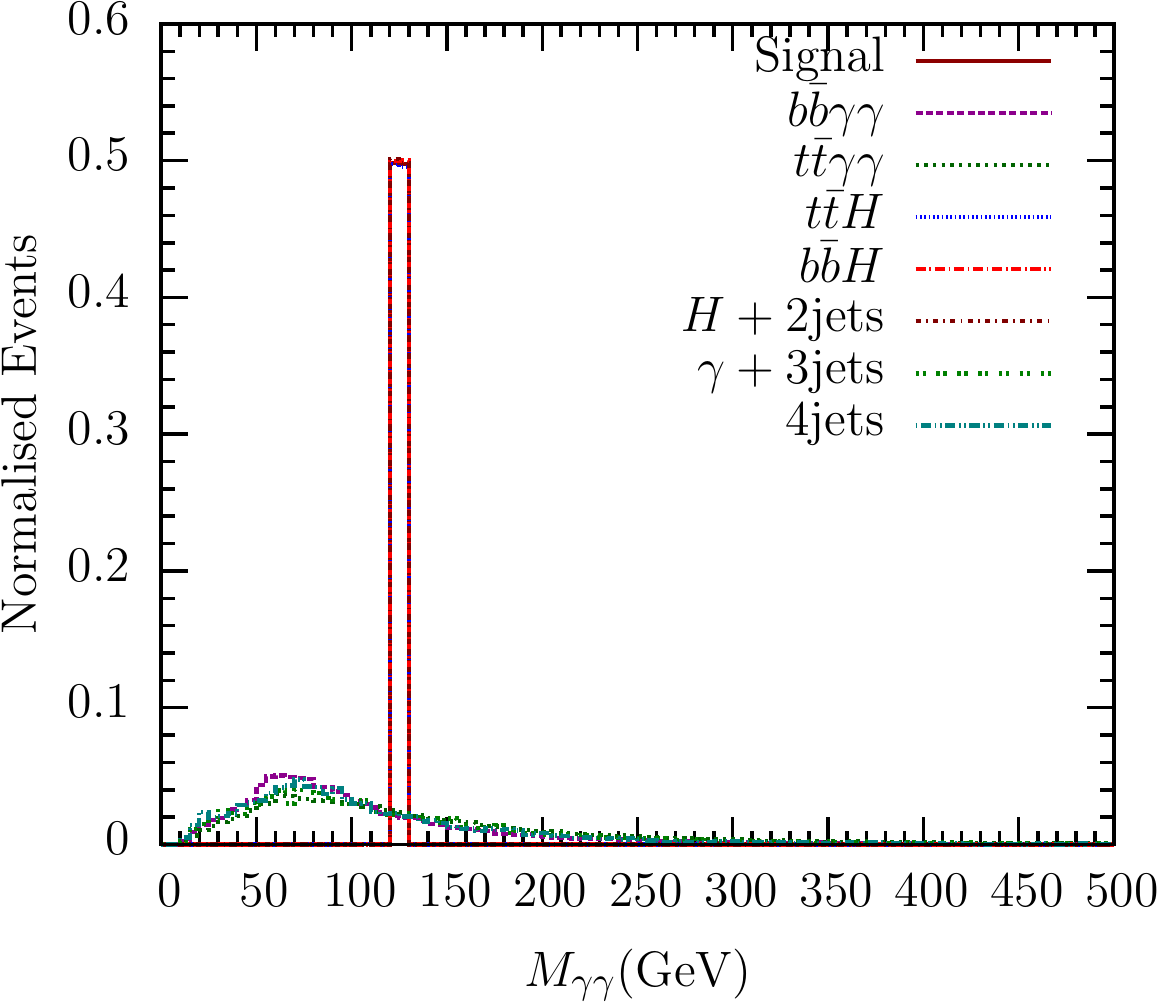}\hspace{5mm}
\includegraphics[scale=0.65]{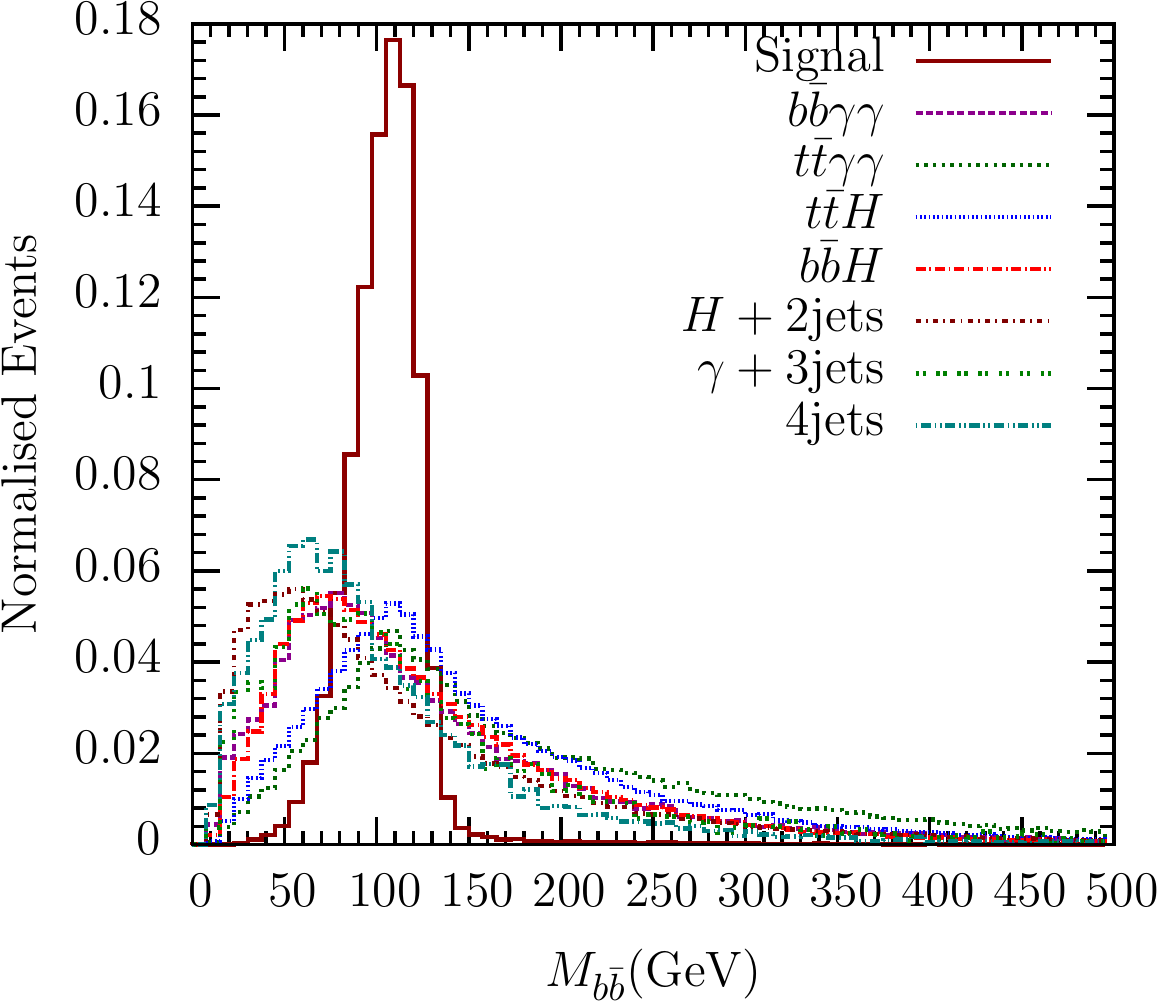}
\caption{\label{bbgg:InvM}Invariant mass distribution for $\gamma\gamma$ and $b\bar b$ pairs in $b\bar b \gamma\gamma\MET$ 
signal for di-Higgs production process at the 14 TeV LHC.}
\end{figure}

\item {\bf Missing transverse Energy, $\MET$ distribution (Cut 4)}: 
In this analysis, the di-Higgs signal arise from the decays of two lightest neutralinos to Higgs and neutrino. The neutrinos coming from heavy particles are expected to have a large transverse momentum contributing to $\MET$. In Fig.~\ref{bbgg:MET}, we show the $\MET$ distribution for the signal as well as backgrounds. For the $M_{\chi^0_1}=300$ GeV, the $\MET$ distribution is expected to peak at around 200 GeV. For the non-top backgrounds, the only source of $\MET$ comes from uncertainties in the measurements of $p_T$ of the various objects at the detector, dominant being jets. Thus, the peaks in $\MET$ distributions for such backgrounds are at very low $\MET$. On the other hand, the backgrounds, which contain top-pairs, produce neutrinos when they decay semi-leptonically and thus may have large $\MET$. From Fig.~\ref{bbgg:MET} we see that  the $\MET$ distribution for a non-top background almost vanish at ($\le 80$ GeV) while the backgrounds containing a top-pair have significant $\MET$ until $\MET<200$ GeV. For the signal, the peak is at 200 GeV and the distribution remains significant until a very large value of $\MET\sim 800$ GeV. Thus, we further put a cut of $$ \MET > 100 \mbox{GeV} $$ to further suppress the backgrounds. With this cut, the $b\bar b\gamma\gamma$ background is completely eliminated and the total background is suppressed by a factor of 30 while the signal events are affected only by 20\%.

\begin{figure}[h!]\centering
\includegraphics[scale=0.7]{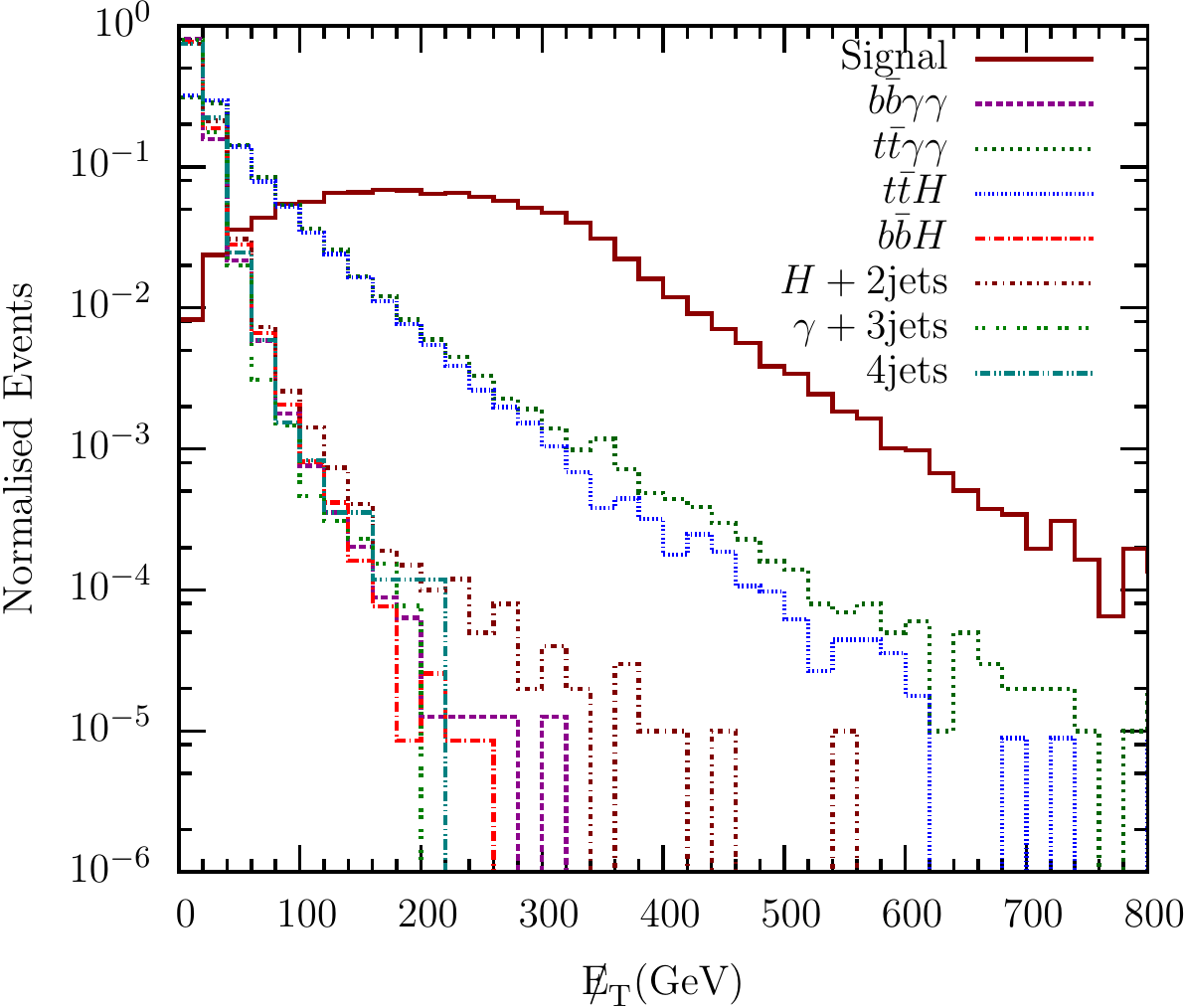}
\caption{\label{bbgg:MET}Missing transverse energy, $\MET$, distribution for $b\bar b \gamma\gamma\MET$ signal for the di-Higgs process and various backgrounds at the 14 TeV LHC.}
\end{figure}
 
\begin{figure}[h!]
\includegraphics[scale=0.65]{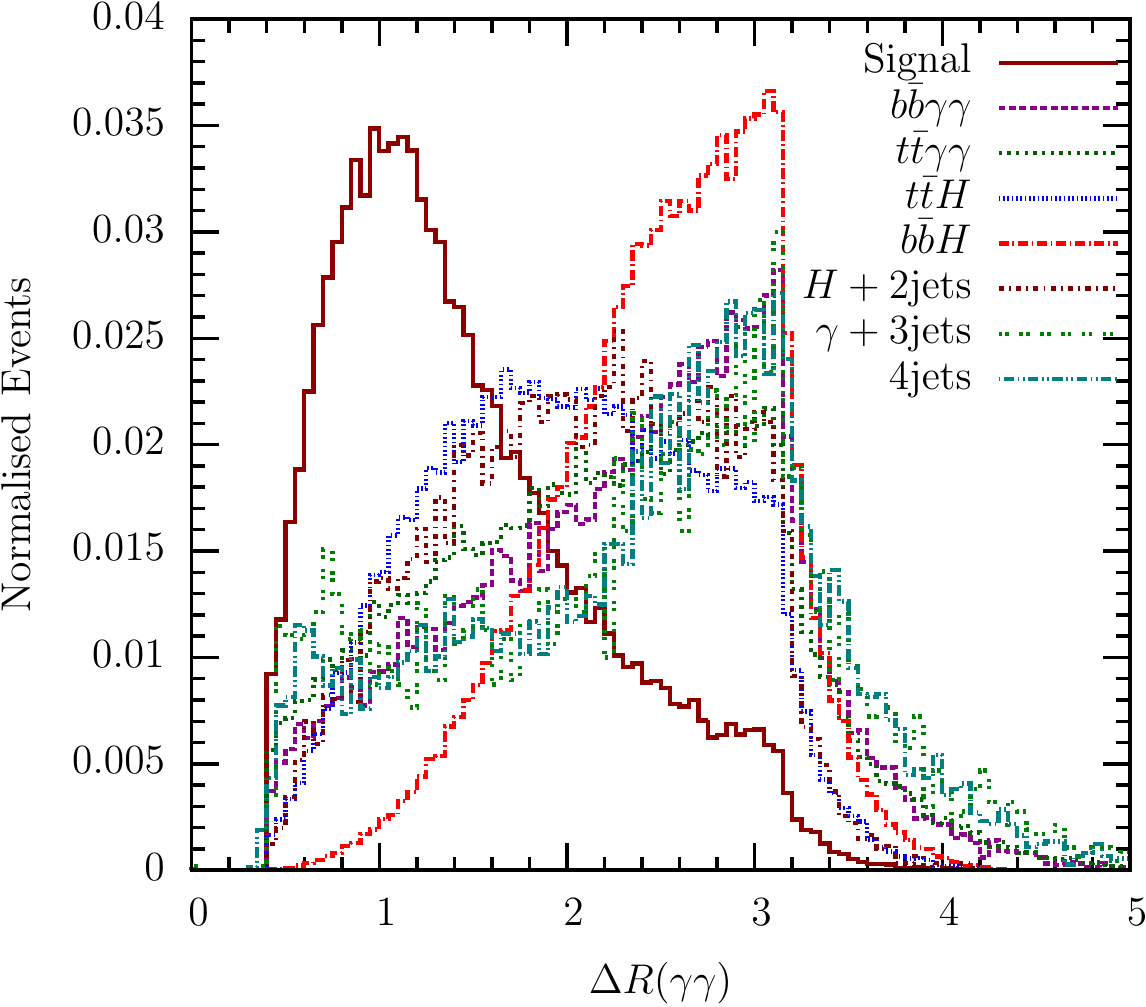}\hspace{5mm}
\includegraphics[scale=0.65]{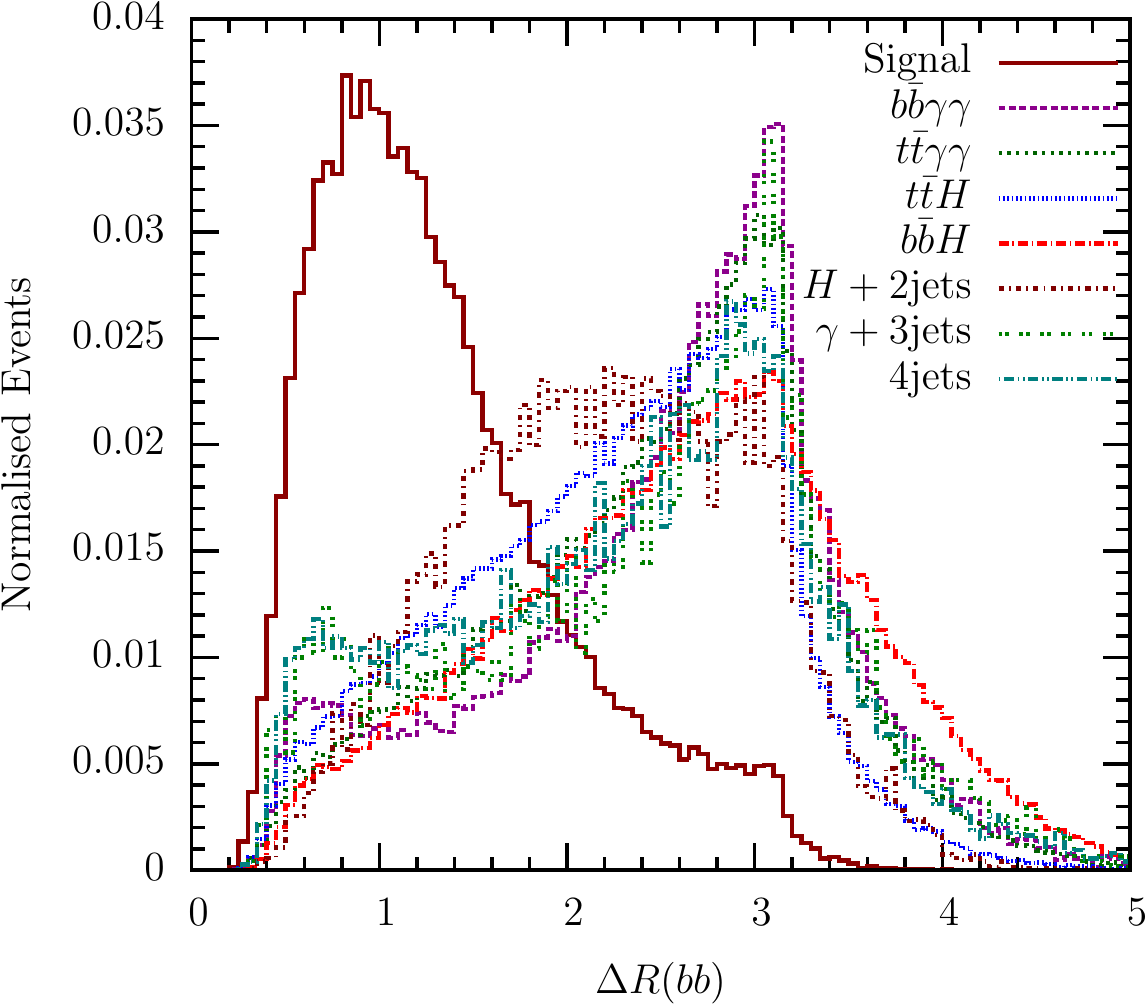}
\caption{\label{bbgg:DelR} $\Delta R$ distribution for $\gamma\gamma$ (left) and $b\bar b$ pairs (right) in $b\bar b \gamma\gamma\MET$ signal for the di-Higgs process and various backgrounds at the 14 TeV LHC.}
\end{figure}
 
 \item {\bf $\Delta R$ separation (Cut 5)}: We find two interesting angular correlations which are significantly different between signal and backgrounds. The $\Delta R$ separation of the  photon-photon and $b\bar b$ pairs are small for the signals and larger for the backgrounds. The shape of the signal distribution stems from the fact that the photon pair and $b$ pair arise from the highly boosted resonant Higgs bosons. These Higgs bosons are boosted as they are produced from the decays of heavy neutralinos/charginos. On the other hand, diphotons and di-$b$ jets for the backgrounds do not come from any heavy resonance and thus are expected to be farther spaced. This fact can be seen from Fig.~\ref{bbgg:DelR} where we show the $\Delta R$ distribution for $\gamma \gamma$ and $b\bar b$ pairs. A cut on $\Delta R$ can effectively suppress the background relative to the signal. Based on these observations, we put following additional cuts: $$\Delta R_{b\bar b} <2.0, ~~~~~~\Delta R_{\gamma\gamma} <2.0.$$ These cuts cuts reduce the background by an order of magnitude while the signal events are reduced by only 20\%.
 \end{itemize}
\end{itemize}

\begin{table}[h!]
\centering
 \newcolumntype{C}[1]{>{\centering\let\newline\\\arraybackslash\hspace{0pt}}m{#1}} 
\begin{tabular}{ |C{2.cm}||C{2.cm}C{2.cm} C{2.cm}  C{2.cm} C{2.cm}|}

 \hline
 \hline
 $\mu$ (GeV) 	& Cut 1 (fb) & Cut 2 (fb) & Cut 3 (fb)	& Cut 4 (fb) & Cut 5 (fb)\\
\hline 	
 $300$    & 9.6$\times 10^{-2}$	& 9.0$\times 10^{-2}$ 	& 5.3$\times 10^{-2}$ 	& 3.8$\times 10^{-2}$	& 1.9$\times 10^{-2}$ \\
 $400$    & 3.5$\times 10^{-2}$	& 3.3$\times 10^{-2}$ 	& 2.0$\times 10^{-2}$ 	& 1.6$\times 10^{-2}$	& 1.1$\times 10^{-2}$ \\
 $500$    & 1.5$\times 10^{-2}$	& 1.4$\times 10^{-2}$ 	& 8.5$\times 10^{-3}$ 	& 7.6$\times 10^{-3}$	& 5.9$\times 10^{-3}$ \\
 $600$    & 7.1$\times 10^{-3}$	& 6.6$\times 10^{-3}$ 	& 4.0$\times 10^{-3}$ 	& 3.7$\times 10^{-3}$	& 3.2$\times 10^{-3}$ \\		
 \hline
 \end{tabular}
\caption{\label{cut-flow:sig} Effects of the cut flow (discussed in Section \ref{cuts}) on the signal events. The $BR(\chi_1^0\to \nu h)$
is assumed to be $100\%$ for all benchmark points.}
\end{table}

\begin{table}[h!]
\centering
 \newcolumntype{C}[1]{>{\centering\let\newline\\\arraybackslash\hspace{0pt}}m{#1}} 
\begin{tabular}{ |C{2.cm}||C{2.cm}C{2.cm} C{2.cm}  C{2.cm} C{2.cm}|}

 \hline
 \hline
			& Cut 1 (fb) & Cut 2 (fb) & Cut 3 (fb)	& Cut 4 (fb) & Cut 5 (fb)\\
\hline 			
 $b\bar b \gamma\gamma$	& 4.3$\times 10^{1}$	& 1.3$\times 10^{1}$	& 4.5$\times 10^{-2}$ 	& 2.1$\times 10^{-4}$	& 1.0$\times 10^{-4}$ \\
 $H b \bar b$		& 9.5$\times 10^{-3}$	& 9.0$\times 10^{-3}$  	& 1.5$\times 10^{-3}$ 	& 1.0$\times 10^{-6}$	& 4.8$\times 10^{-7}$\\
 $H j j$		& 2.9$\times 10^{-5}$	& 2.8$\times 10^{-5}$ 	& 5.5$\times 10^{-6}$ 	& 1.1$\times 10^{-8}$	& 1.0$\times 10^{-8}$\\
 $t\bar t\gamma\gamma$	& 1.2$\times 10^{0}$	& 6.1$\times 10^{-1}$ 	& 2.2$\times 10^{-3}$ 	& 2.5$\times 10^{-4}$	& 4.7$\times 10^{-5}$\\
 $t\bar tH$		& 1.1$\times 10^{-1}$	& 1.0$\times 10^{-1}$ 	& 2.0$\times 10^{-2}$ 	& 1.9$\times 10^{-3}$	& 5.0$\times 10^{-4}$\\
 $b\bar{b}jj$		& 4.2$\times 10^{1}$	& 3.5$\times 10^{1}$ 	& 1.5$\times 10^{-1}$	& 1.6$\times 10^{-3}$ 	& 4.0$\times 10^{-4}$\\
  $jj\gamma\gamma$	& 9.3$\times 10^{-2}$ 	& 2.6$\times 10^{-2}$	& 8.9$\times 10^{-5}$	& $-$ 	&  $-$\\
 $jjjj$			& 1.8$\times 10^{-2}$	& 1.5$\times 10^{-2}$	& 5.6$\times 10^{-5}$ 	& $-$	&  $-$ \\
 \hline
 $\Sigma$ (bckg.)	& 	&  	&  	& 	& 1.1$\times 10^{-3}$ \\
 \hline\hline
 \end{tabular}
\caption{\label{cut-flow:bbgg} Effects of the cut flow on the background events. }
\end{table}

In Table \ref{cut-flow:sig}, we show the cut flow of the cross sections for the signal for different values of the Higgsino masses. One can observe that the cross section rapidly decreases with the increase in masses but the efficiency of the cuts is better for the heavier Higgsinos. In Table \ref{cut-flow:bbgg}, we show the cut flow of the cross sections for the various background processes  considered in the analysis. One can see from Table \ref{cut-flow:bbgg}, the dominant contributions to the background come from the $b\bar b \gamma\gamma$ continuum, $t\bar t \gamma \gamma$ and $t\bar t h$ processes, each having cross sections of 44 fb, 1.2 fb and 0.11 fb respectively after cut 1. On the other hand, the signal cross section after cut 1 is 0.096 fb for $M_{\tilde \chi^0}=300$ GeV. Thus the signal-to-background ratio at this stage of cut flow is only $5\times10^{-4}$. The cut on missing transverse energy $\MET>100$ GeV (cut 4) almost eliminates the $b\bar b\gamma\gamma$ background while the contributions of $t\bar t\gamma\gamma$ and $t\bar t h$ are reduced to $10^{-4}$ and $10^{-3}$ respectively. After the cut 4, the $S/B$ ratio is 6 which is a tremendous improvement over previous value after cut 1. The cut 5 (on $\Delta R$) further suppresses the background contribution and thus enhancing the $S/B$ ratio. We estimate the statistical significance of the signal using following formula \cite{CMS-Sig}

\begin{equation}\label{Eq:sig}
Sig=\sqrt{2((S+B)ln(1+\frac{S}{B})-S)}
\end{equation}
where $S$ and $B$ are the number of signal and background events, respectively.

We will now comment on the possibility of detecting the di-Higgs signal for different masses of the Higgsinos. In Table \ref{cut-eff}, we show the cut efficiency of the signal for different masses of Higgsinos.  As the masses increase, the total production cross section goes down rapidly, but the handle over the background improves significantly. Thus, even though the cross section for the signal is decreased with the mass, the efficiency for the detection of the signal is increased. 

\begin{table}[h!]
\centering
 \newcolumntype{C}[1]{>{\centering\let\newline\\\arraybackslash\hspace{0pt}}m{#1}} 
\begin{tabular}{ |C{2.5cm}||C{3.5cm}|}
 \hline
 \hline
 $\mu$ (GeV) 	& Cut efficiency (\%)\\
\hline 	
 $200$    &    	1.32 \\
 $300$    &    	4.48 \\
 $400$    &    	6.76 \\
 $500$    &    	8.50 \\
 $600$    &  	9.62 \\		
 \hline
 \hline
 \end{tabular}
\caption{\label{cut-eff} The cut efficiencies for the signal for different Higgsino masses at the 14 TeV LHC.}
\end{table}

\begin{figure}[h!]\centering
\includegraphics[scale=0.75]{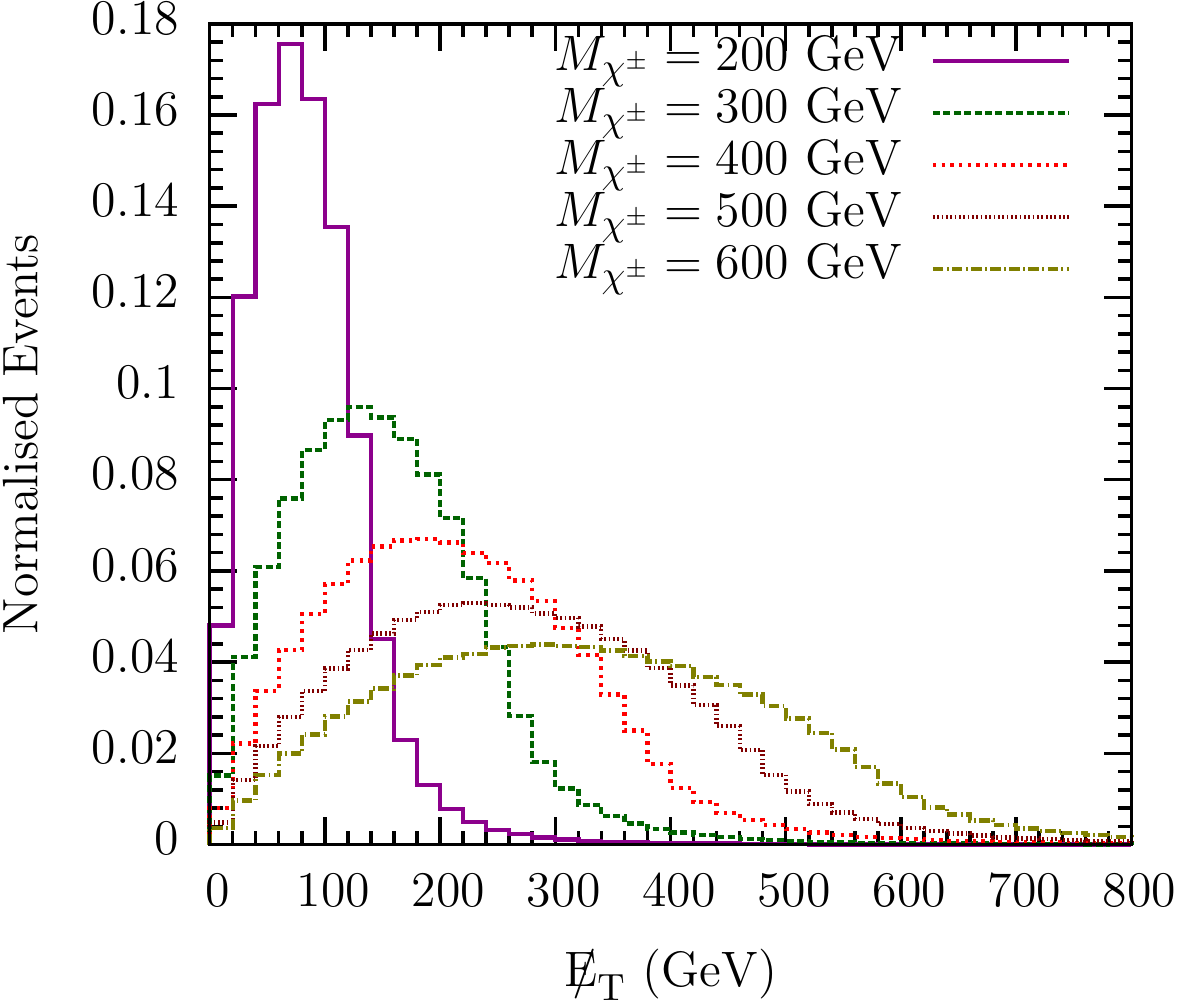}
\caption{\label{bbgg:MET_mchi} The $\MET$ distribution for the di-Higgs process for different values of chargino masses at the 14 TeV LHC.}
\end{figure}

Let us now look at the various distributions discussed earlier, for the different values of Higgsino masses. In Fig.~\ref{bbgg:MET_mchi}, we show the missing transverse energy ($\MET$) distribution for the di-Higgs signal for different values of the Higgsino masses ranging from 200 GeV to 600 GeV. We notice that the $\MET$ distribution peaks at 60 GeV, 100 GeV, 150 GeV, 200 GeV and 250 GeV for $M_{\chi^\pm}=$ 200 GeV, 300 GeV, 400 GeV, 500 GeV and 600 GeV, respectively. Thus, in order to further improve the cut efficiency, optimizing of the cuts on $\MET$ should be performed for different Higgsino masses. However, one should keep in mind that as the backgrounds are already low, any further stringent cuts would also result in reducing the signal events.

\begin{figure}[h!]\hspace{-2.5mm}
\includegraphics[scale=0.7]{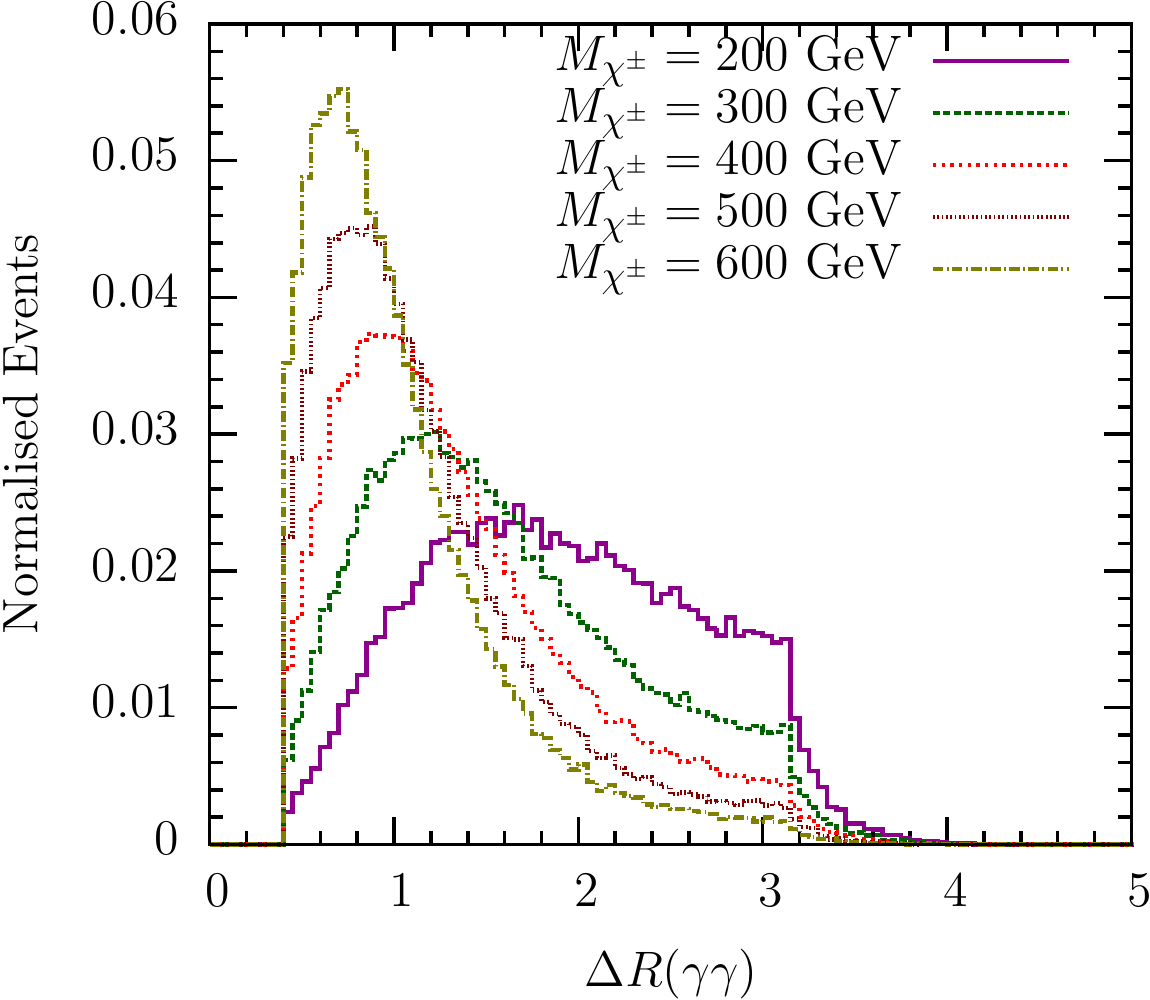}\hspace{5mm}
\includegraphics[scale=0.7]{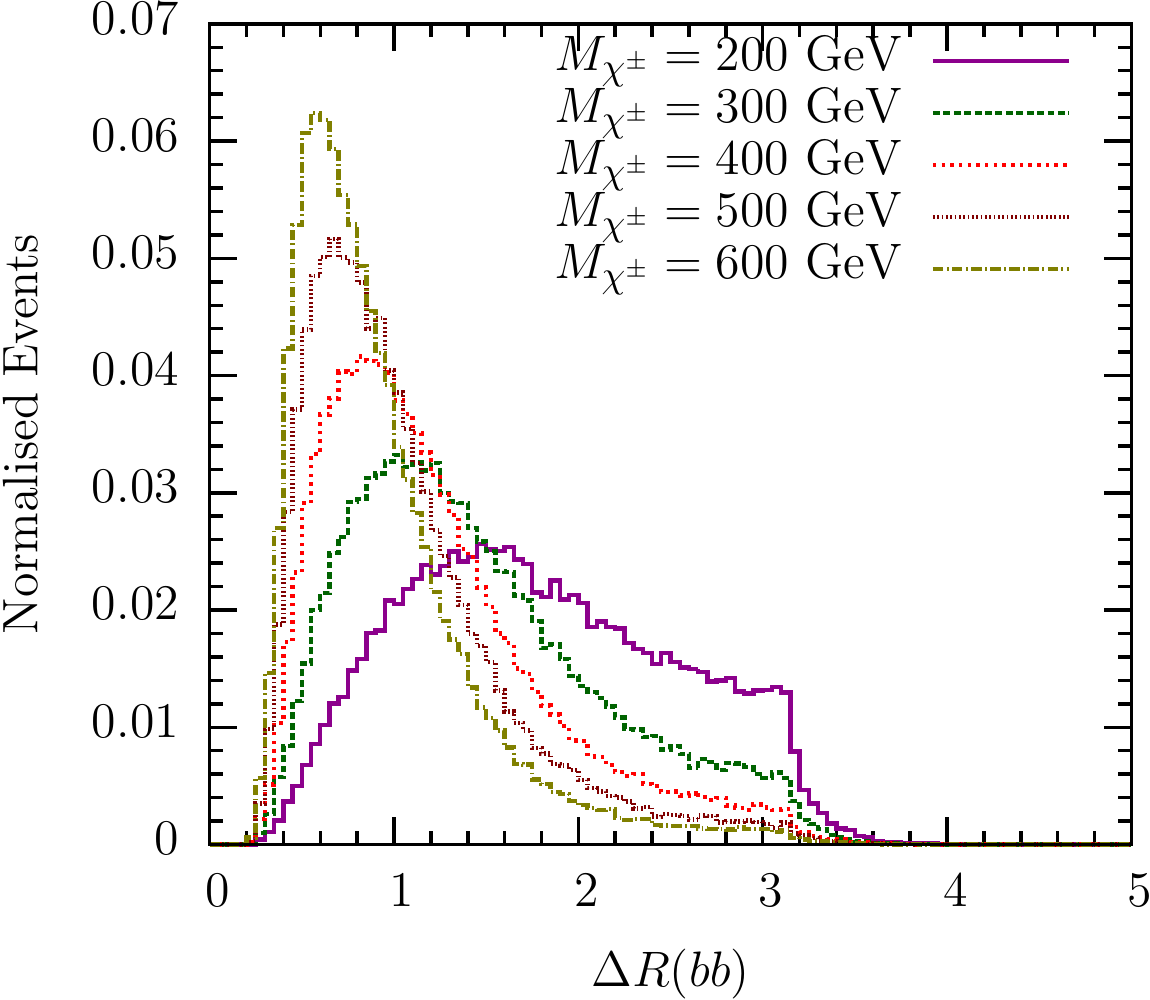}
\caption{\label{bbgg:DelR_mchi} The $\Delta R$ distribution for $\gamma\gamma$ (left) and $b\bar b$ pairs (right) for the di-Higgs process for different values of chargino masses at the 14 TeV LHC.}
\end{figure}

In Fig.~\ref{bbgg:DelR_mchi}, the $\Delta R$ separation between the photon-photon (left) and $b\bar b$ pairs (right) for the di-Higgs+$\MET$ signal at the 14 TeV LHC have been shown. For  very heavy Higgsinos, the Higgs coming from the decay will be highly boosted. Thus the $\Delta R$ separation between $b\bar b$/ $\gamma\gamma$ becomes very small. This fact determines the shape of the $\Delta R$ distributions for different chargino masses. For the mass of 600 GeV, the distribution is peaked at a very low value of $\Delta R$ while for a light chargino of 200 GeV, the diphotons and di-$b$ jets are much farther spaced. Based on these findings, the cuts on $\Delta R$ need to be judiciously chosen so as to enhance the signal-to-background ratio for the different values of Higgsino masses. 

\begin{table}[h!]
\centering
 \newcolumntype{C}[1]{>{\centering\let\newline\\\arraybackslash\hspace{0pt}}m{#1}} 
\begin{tabular}{ |C{2.5cm}||C{3.cm}C{3.cm}C{3.cm}|}
 \hline
 \hline
 $\mu$ (GeV) 	& Sig. (1 ab$^{-1}$) & Sig.  (2 ab$^{-1}$)& Sig.  (3 ab$^{-1}$)\\
\hline 	
 $300$    &    	10.3 &  14.6	& 17.2 \\
 $400$    &    	6.6  &  9.3	& 11.3\\
 $500$    &    	4.1  &  5.7	& 7.1\\
 $600$    &  	2.4  &  4.2	& 5.7\\		
 \hline
 \hline
 \end{tabular}
\caption{\label{tab:sig} The signal significance for different values of $\mu$ for three different integrated luminosities of 1, 2 and 3 ab$^{-1}$ at the 14 TeV LHC.}
\end{table}

\begin{figure}[h!]\hspace{-3.5mm}
\includegraphics[scale=0.65]{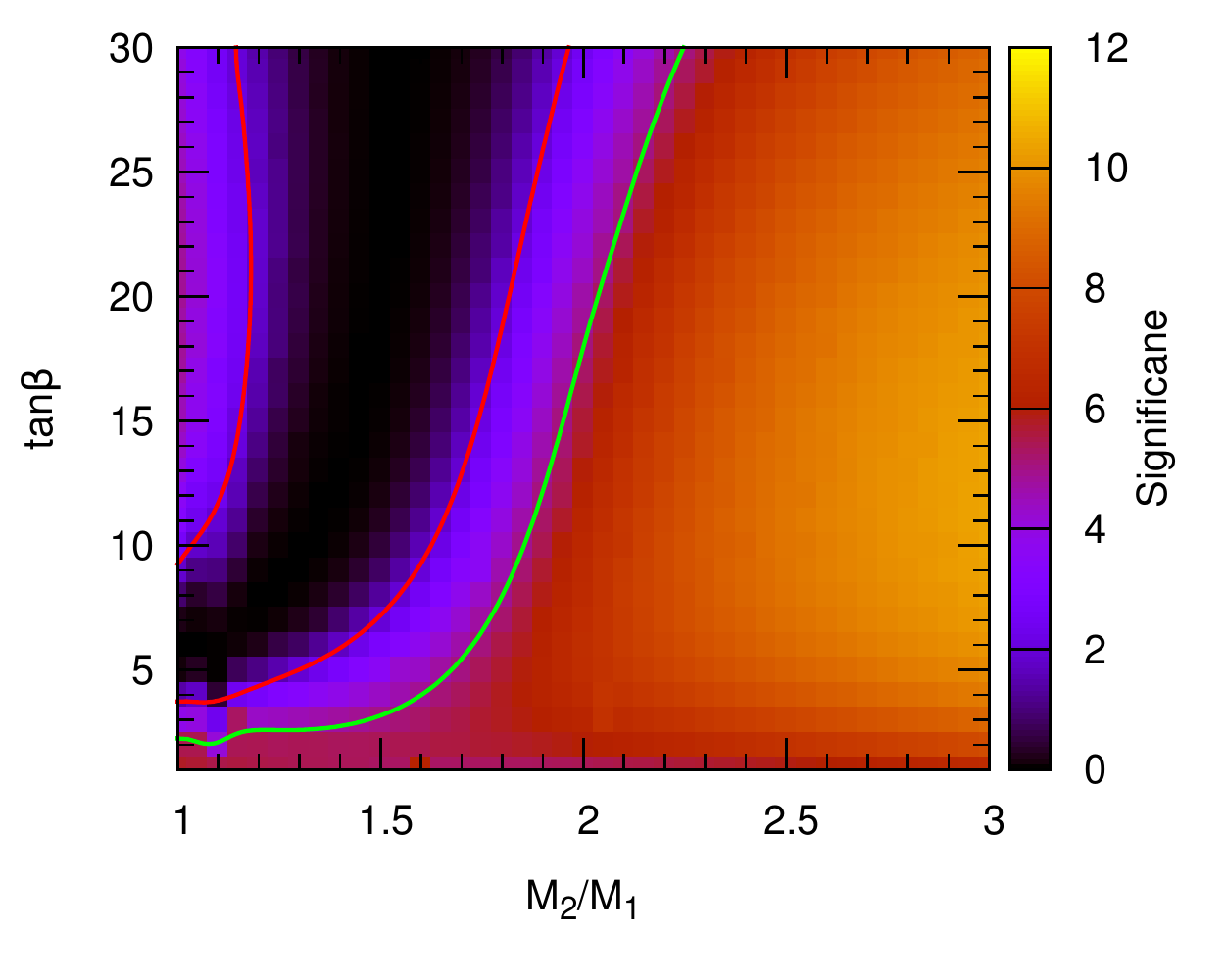}
\includegraphics[scale=0.65]{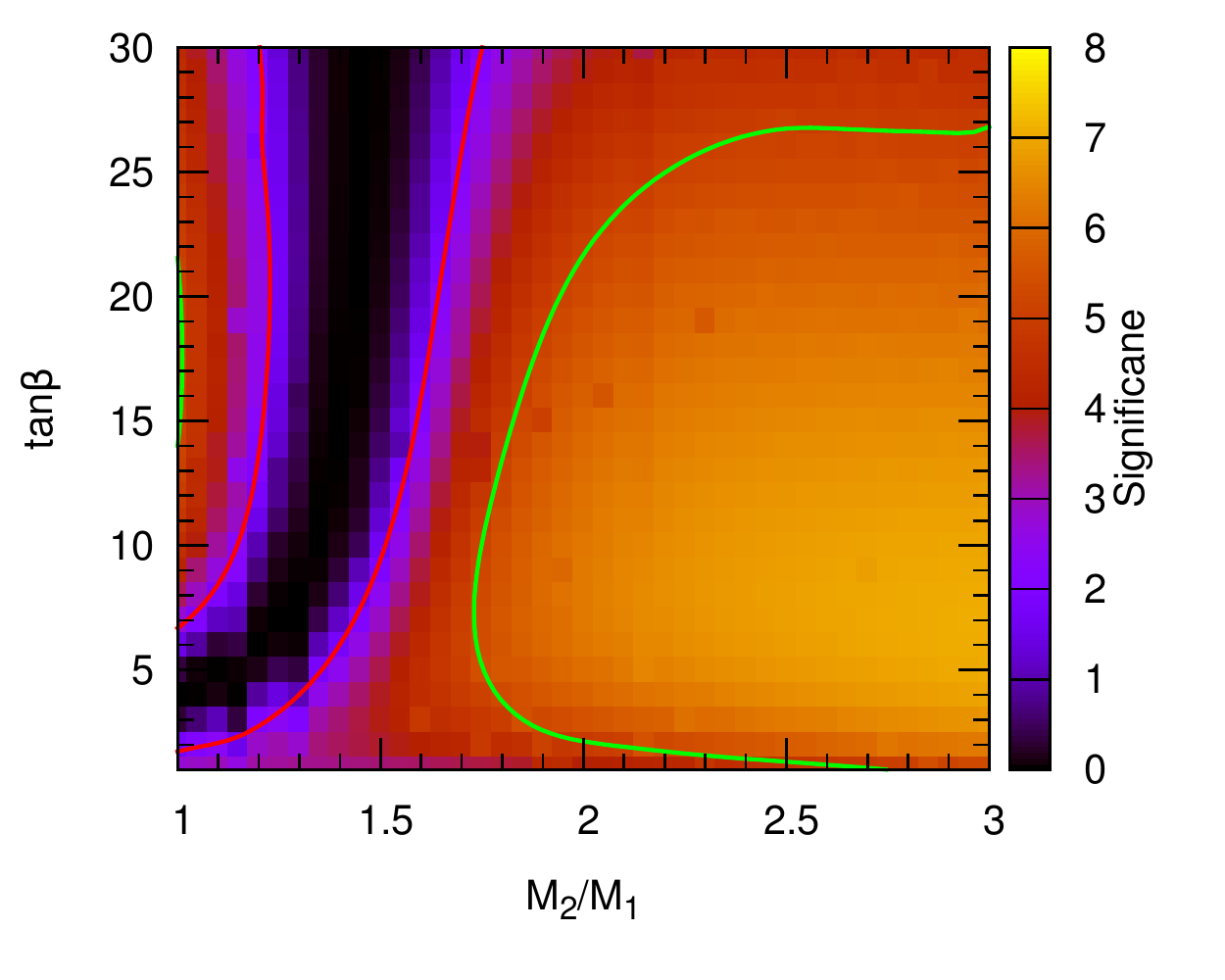}
\caption{\label{fig:sig} The signal significance in the $(M_2,\tan\beta)$ plane for two values of $\mu=$ 300 GeV (left) and 500 GeV (right) at the 14 TeV LHC. For $\mu=$ 300 GeV (500 GeV), integrated luminosities of 1 ab$^{-1}$ (3 ab$^{-1}$) has been taken for the analysis. The region outside the red and green curves show the regions of the $2\sigma$ exclusion and the $5\sigma$ discovery,
respectively.}
\end{figure}

We now discuss the exclusion/discovery limits for the Higgsino LSP in di-Higgs production at the LHC14 for the different values of integrated luminosities. Based on the cut-flow analysis for the signal and backgrounds, we estimate the significance of the signal, using Eq.~\ref{Eq:sig}, for various values of $\mu$ and for three different values of integrated luminosities 1, 2 and 3 ab$^{-1}$. 
 The numbers for the significance have been presented in  Table \ref{tab:sig}. For a light LSP of $M_{\tilde\chi^0_1}\sim 400$ GeV, it would be possible to discover them with more than 5$\sigma$ significance at 1 ab$^{-1}$ of integrated luminosity. On the other hand, for a heavy LSP $M_{\tilde\chi^0_1}\sim 600$ GeV, it will need around 3 ab$^{-1}$ of data to have sufficient discovery prospects. 

In Fig.~\ref{fig:sig} we show the statistical significance for the signal in the plane of $(M_2,\tan\beta)$ for two different values of $\mu=300$ (left) and 500 GeV (right). As mentioned earlier, Br($\tilde\chi^0_1\to\nu h$) also depends on parameters $\tan\beta$ and $M_2$ other than $\mu$. The partial width in individual decay mode of the LSP depends on ratio of $M_2/M_1$ and $\tan\beta$ on such a way that the branching ratio to $\nu h$ is minimum when $M_2/ M_1 \approx 1$. Strictly speaking this depends on $\tan\beta$ as well. This is also evident from Fig. \ref{fig:mu-m2}. The dark regions in the figures denote the regions for low discovery prospects while the brighter region has better prospects to observe a Higgsino LSP in di-Higgs process. For $\mu=300$ GeV, we find that it would be possible to discover the Higgsino-like LSP in di-Higgs production process at the LHC with only 300 fb$^{-1}$ of integrated luminosity and a large portion of the parameter can be excluded at 2$\sigma$ with 1 ab$^{-1}$. The region of exclusion has been shown by red curves in the figures. The 5$\sigma$ discovery prospect has been denoted by green curves in both the plots. For $\mu=500$ GeV, the 5$\sigma$ discovery can only be obtained with full data set of the 14 TeV LHC {\it viz.} 3 ab$^{-1}$ for large values of $M_2$.

\section{Summary and Conclusions}
Motivated by the naturalness, we study a Bilinear RpV SUSY scenario where the LSP is Higgsino-like, and BRpV couplings determine the tree-level neutrino mass matrix. We investigate the parameter space of this scenario and study the decay patterns of the Higgsino LSP. We find that in a large part of the parameter space, the LSP decays to $\nu h$ with branching fraction larger than 0.9. A large region of this yet unexplored parameter space can still be insensitive to the existing constraints coming from the LHC searches. We then study the pair production of the electroweakinos followed by the decays $\tilde\chi^\pm_1\to \tilde\chi^0_1 W^{\pm*}$ and $\tilde\chi^0_1\to \nu h$. This leads to a very interesting signature of Higgs boson pair production at the LHC accompanied with significant missing transverse energy. This di-Higgs production in BRpV model, occurring in association with a large missing transverse energy, is in distinct contrast to the Higgs boson pair production in the SM. This fact makes this signal quite feasible to search at the luminosity improved version of the 14 TeV LHC despite having a very small cross section.

Among the various decay channels for di-Higgs, we focus on the scenario where one of the Higgses decays to diphoton and the other decays to a bottom pair. This particular decay has the advantage of manageable  SM backgrounds. Thus the signal which we are looking for includes 2 photons, 2 $b$ jets and large missing transverse energy. We perform a realistic detector level simulation for the signal $\gamma\gamma b\bar b+\MET$ taking some benchmark points in the parameter space at the 14 TeV LHC. We also perform a full systematic study of all the background processes. It is found that the cuts on $\MET$ and $\Delta R$ are instrumental in eliminating the QCD multijet backgrounds and suppressing the total background. We also notice that even though the cross sections for the signal decrease as the masses of the Higgsinos get heavy, the increased efficiency of the $\MET$ and $\Delta R$ cuts helps to compensate the overall signal to background ratio. 
Finally we conclude that the LSP of mass 300-500 GeV would be amenable to discovery in the early LHC14 data in the di-Higgs channel. On the other hand for the heavy LSP $\sim$ 600 GeV, it would require full data set (3 ab$^{-1}$) of 14 TeV run to have reasonable discovery prospects.

\medskip
{\bf Acknowledgment}: 
EJC is supported by the NRF grant funded by the Korea government (MSIP)
(No. 2009-0083526) through KNRC at Seoul National University. 

\section*{Appendix A: Effective R-parity violating vertices}
\underline{Neutrino-neutralino diagonalization}

Rotating away the neutrino-neutralino  mixing mass terms 
(by $\theta^N$) can be
made by the following redefinition of  neutrinos and neutralinos:
\begin{equation} 
 \pmatrix{ \nu_i \cr \chi^0_j } \longrightarrow
 \pmatrix{ \nu_i- \theta^N_{ik} \chi^0_k  \cr
           \chi^0_j + \theta^N_{lj} \nu_l }
\end{equation}
where $(\nu_i)$ and $(\chi^0_j)$ represent three neutrinos
$(\nu_e, \nu_\mu, \nu_\tau)$ and four neutralinos
$(\tilde{B}, \tilde{W}_3, \tilde{H}^0_1, \tilde{H}^0_2)$ 
in the flavor basis, respectively.  The rotation elements
$\theta^N_{ij}$ are given by 
\begin{eqnarray} \label{thetaN}
 \theta^N_{ij} &=& \xi_i c^N_j c_\beta - \epsilon_i \delta_{j3} 
 \quad\mbox{and} \\
 (c^N_j) &=& {M_Z \over F_N} ({ s_W M_2 \over c_W^2 M_1 + s_W^2 M_2},
  -{ c_W M_1 \over c_W^2 M_1 + s_W^2 M_2}, -s_\beta{M_Z\over \mu},
   c_\beta{M_Z\over \mu}) \nonumber
\end{eqnarray}
where  $F_N=M_1 M_2 /( c_W^2 M_1 + s_W^2 M_2) + M_Z^2 s_{2\beta}/\mu$.
Here  $s_W=\sin\theta_W$ and $c_W=\cos\theta_W$ 
with the weak mixing angle $\theta_W$.

\underline{Charged lepton/chargino diagonalization}

Defining  $\theta^L$ and $\theta^R$  as the two rotation matrices
corresponding to the left-handed negatively  and positively charged 
fermions, we have
\begin{equation}
 \pmatrix{ e_i \cr \chi^-_j } \rightarrow
 \pmatrix{ e_i- \theta^L_{ik} \chi^-_k  \cr
           \chi^-_j + \theta^L_{lj} e_l } \quad;\quad
 \pmatrix{ e^c_i \cr \chi^+_j } \rightarrow
 \pmatrix{ e^c_i- \theta^R_{ik} \chi^+_k  \cr
           \chi^+_j + \theta^R_{lj} e^c_l } 
\end{equation}
where $e_i$ and $e^c_i$ denote the left-handed charged leptons and
anti-leptons, $(\chi^-_j)=(\tilde{W}^-,\tilde{H}^-_1)$ and 
$(\chi^+_j)=(\tilde{W}^+,\tilde{H}^+_2)$.
The rotation elements $\theta^{L,R}_{ij}$ are  given by
\begin{eqnarray}\label{cR}
&& \theta^L_{ij}= \xi_i c^L_j c_\beta-\epsilon_i \delta_{j2}\;, \quad
 \theta^R_{ij}= {m^e_i\over F_C} \xi_i c^R_j c_\beta  \quad\mbox{and} \\
&&  (c^L_j)= -{M_W \over F_C} (\sqrt{2}, 2s_\beta{M_W\over \mu})\;,
               \nonumber \\
&&   (c^R_j)= -{M_W  \over F_C} (\sqrt{2}(1-{M_2\over \mu} t_\beta), 
      \frac{M_2^2 c^{-1}_\beta}{\mu M_W }+2{M_W \over \mu} c_\beta) 
      \nonumber
\end{eqnarray}
and $F_C= M_2 + M_W^2 s_{2\beta}/\mu$.

\underline{Sneutrino/neutral Higgs boson diagonalization}

Denoting the rotation matrix by $\theta^S_i=a_i$, we get
\begin{equation} \label{thetaS}
 \pmatrix{ \tilde{\nu}_i \cr H^0_1 \cr} \rightarrow
 \pmatrix{ \tilde{\nu}_i + a_i H^0_1   \cr
     H^0_1 - a_i \tilde{\nu}_i \cr}
\end{equation}

\medskip

With the expressions for the rotation matrices, we can obtain the effective R-parity violating vertices from the usual R-parity conserving interaction vertices, which are relevant to the LSP decays. We list them below by taking only the linear terms in $\theta$'s which are enough for our purpose.

 
\underline{$\chi^0-l-W$ vertices}:
\begin{eqnarray} \label{Vert:chilW}
 {\cal L}_{\chi^0 l W} &=& \overline{\chi^0_i}\gamma^\mu 
  \left[ P_L L^{\chi^0 l W}_{ij}  + P_R R^{\chi^0 l W}_{ij} \right]
          e_j W_\mu^+ + h.c. 
\\ 
\mbox{with} \quad L^{\chi^0 l W}_{ij} &= &{g\over \sqrt{2}} \,
             [c^N_1,c^N_2-\sqrt{2}c^L_1, c^N_3-c^L_2,c^N_4] \,
             \xi_j c_\beta \nonumber \\
           \quad R^{\chi^0 l W}_{ij} &= &{g\over \sqrt{2}} \,
             [0,-\sqrt{2}c^R_1, 0, -c^R_2] \, {m^e_j \over F_C} 
             \xi_j c_\beta 
\nonumber
\end{eqnarray}

\underline{$\chi^0-\nu-Z$ vertices}:
\begin{eqnarray} \label{Vert:chinuZ}
 {\cal L}_{\chi^0 \nu Z} &=& \overline{\chi^0_i}\gamma^\mu P_L 
        L^{\chi^0 \nu Z}_{ij} \nu_j Z_\mu^0 + h.c. 
\\ 
\mbox{with} \quad L^{\chi^0 \nu Z}_{ij} &= &{g\over 2 c_W}\,
             [c^N_1,c^N_2,0,2c^N_4]\;\xi_j c_\beta \,.
\nonumber
\end{eqnarray}

\underline{$\chi^0-\nu-h$ vertices}:
\begin{eqnarray}\label{Vert:chinuh}
 {\cal L}_{\chi^0 \nu h} &=& \overline{\chi^0_i} 
   P_L L^{\chi^0 \nu h}_{ij}  \nu_j\, h + h.c. 
\\
\mbox{with} \quad 
  L^{\chi^0 \nu h}_{ij} &= & {g\over2 c_W} \,
             [s_W (1-c^N_{3} c_\beta + c^N_4 s_\beta), 
             -c_W  (1 - c^N_{3} c_\beta + c^N_4 s_\beta), 
\nonumber\\
 && ~~~ (s_W c^N_{1}- c_W c^N_{2}) c_\beta, (s_W c^N_1 - c_W c^N_2 ) s_\beta] \xi_j c_\beta
             \nonumber 
\end{eqnarray}

 
\underline{$\chi^+-\nu-W$ vertices}:
\begin{eqnarray} \label{Vert:chanuW}
 {\cal L}_{\chi^+ \nu  W} &=& \overline{\chi^-_i}\gamma^\mu \left[
  P_L L^{\chi^+ \nu W}_{ij} + P_R  R^{\chi^+ \nu W}_{ij} \right] 
  \nu_j W_\mu^- + h.c. 
\\ 
\mbox{with} \quad L^{\chi^+ \nu W}_{ij} &= &{g\over \sqrt{2}} \,
             [c^L_1 - \sqrt{2} c^N_2, c^L_2-c^N_3] \,
             \xi_j c_\beta \nonumber \\
           \quad R^{\chi^+ \nu  W}_{ij} &= &{g\over \sqrt{2}} \,
             [-\sqrt{2}c^N_2, c^N_4] \,
             \xi_j c_\beta 
\nonumber
\end{eqnarray}
  
 \underline{$\chi^+ -l-Z$ vertices}:
\begin{eqnarray} \label{Vert:chalZ}
 {\cal L}_{\chi^+ l Z} &=& \overline{\chi^-_i} \gamma^\mu \left[
  P_L  L^{\chi^+ l Z}_{ij} + P_R R^{\chi^+ l Z}_{ij} \right]
   e_j Z_\mu^0 + h.c. 
\\ 
\mbox{with} \quad L^{\chi^+ l Z}_{ij} &= &{g\over 2 c_W} \,
             [c^L_1,0]\;\xi_j c_\beta \,,
\nonumber\\
\quad R^{\chi^+ l Z}_{ij} &= &{g\over 2 c_W}\,
             [2  c^R_1,c^R_2]\;{m^e_j \over F_C} \xi_j c_\beta \,,
\nonumber
\end{eqnarray}

\underline{$\chi^+ -l-h$ vertices}:
\begin{eqnarray}\label{Vert:chalh}
 {\cal L}_{\chi^+ l h} &=& \overline{\chi_i^-}
  \left[ P_L L^{\chi^+ l h }_{ij}
         + P_R R^{\chi^+ l h }_{ij} \right]
          e_j h + h.c. 
\\
\mbox{with} \quad 
  L^{\chi^+ l h}_{ij} &= &{ \sqrt{2}g} \,
             [c^L_2 c_\beta +1, c^L_1 s_\beta] \xi_j c_\beta
             \nonumber \\
  R^{\chi^+ l h}_{ij} &= &
           [\sqrt{2} g c^R_2 s_\beta - c^L_1 c_\beta {F_C \over v c_\beta}, 
            \sqrt{2} g c^R_1 c_\beta - (c^L_2  c_\beta+1)  {F_C \over v c_\beta}]
          {m^e_j \over F_C} \xi_j c_\beta\,. 
\nonumber
\end{eqnarray}

\section*{Appendix B: Decay widths of neutralinos}
For a generic decay process $\tilde\chi_i \to L_j V$ where L is either $\nu$ or $\ell^\pm$ and $V$ is either $Z$ or $W^\pm$, the decay width can be written as:

\begin{equation}
 \Gamma(\tilde\chi_i \to L_j V) = \frac{G_F \ m_\chi^3}{4\sqrt{2}\pi}\left[\left|C_i^L\right|^2+\left|C_i^R\right|^2\right]|\xi_j|^2 \ c_\beta^2 \ I_2(r_V)
\end{equation}
where $C_i^L$ and $C_i^R$ are the left- and right-handed couplings, $r_V$ is $(m_V^2/m_{\tilde\chi}^2)$ and $I_2(r_V)=(1-r_V)^2\times(1+2 \ r_V)$.

\begin{enumerate}
 
 \item For $\tilde\chi_i^0 \to \nu_j Z$
 \begin{eqnarray}\label{width:vZ}
  C_i^L &=& \frac{1}{2}\left[N_{i1}c_1^N+N_{i2}c_2^N+N_{i4}\ 2c_4^N\right]\nonumber\\
  C_i^R &=& 0
 \end{eqnarray}

 \item For $\tilde\chi_i^0 \to \ell_j W$
 \begin{eqnarray}\label{width:lW}
  C_i^L &=& \frac{1}{\sqrt{2}}\left[N_{i1}c_1^N+N_{i2}(c_2^N-\sqrt{2}c_1^L)+N_{i3}(c_3^N-c_2^L)+N_{i4}\ c_4^N\right]\nonumber\\
  C_i^R &=& \left[9N_{i2}c_1^R + \frac{N_{i4}}{2}\ c_2^R\right] \ \frac{m_j^e}{F_C}
\end{eqnarray}
\end{enumerate}
 
Similarly, for the decay $\tilde\chi_i^0 \to \nu_j  h$, the decay width can be written as
\begin{equation}
\Gamma(\tilde\chi_i^0 \to \nu_j h) = \frac{G_F \ m_\chi \ m_W^2}{4\sqrt{2}\pi}\left[\left|C_i^L\right|^2+\left|C_i^R\right|^2\right]|\xi_j|^2 \ c_\beta^2 (1-r_h)^2
\end{equation}
where $r_h$ is $m_h^2/m_{\tilde\chi}^2$.
 \begin{eqnarray}\label{width:vh}\nonumber
   C_i^L &=& N_{i1}t_W(1-c_3^Nc_\beta+c_4^Ns_\beta)+N_{i2}(-1+c_3^Nc_\beta-c_4^N \ s_\beta)\\\nonumber
         &+& N_{i3}(t_W c_1^N - c_2^N) c_\beta + N_{i4}(t_W c_1^N-c_2^N)s_\beta\\
   C_i^R &=& 0
 \end{eqnarray}

Here, $N$ is the $4\times 4$ matrix and diagonalize the neutralino mass matrices.

\end{document}